\documentclass[%
 aip,
 amsmath,amssymb,
 reprint,%
]{revtex4-1}

\usepackage{graphicx}
\usepackage{subcaption} 
\usepackage{dcolumn}
\usepackage{bm}

\usepackage[utf8]{inputenc}
\usepackage[T1]{fontenc}
\usepackage{mathptmx}
\usepackage{etoolbox}   
\usepackage[section]{placeins}
\usepackage{float}

\makeatletter
\def\@email#1#2{%
 \endgroup
 \patchcmd{\titleblock@produce}
  {\frontmatter@RRAPformat}
  {\frontmatter@RRAPformat{\produce@RRAP{*#1\href{mailto:#2}{#2}}}\frontmatter@RRAPformat}
  {}{}
}%
\makeatother
\begin{document}

\preprint{AIP/123-QED}

\title[Exploratory Study of Chaotic Behavior in Walking Droplets]{Exploratory Study of Chaotic Behavior in Walking Droplets:\\Single and Double Slit Experiments in the Supercritical Faraday Regime}
\author{E. Dunn}
 \email{emily.dunn@duke.edu}
\affiliation{Department of Mechanical Engineering and Materials Science, Duke University.}
\affiliation{Department of Physics, Duke University.}
 
\author{B. Keshavarz}%
\affiliation{Department of Mechanical Engineering and Materials Science, Duke University.}

\author{E. Dowell}
\affiliation{Department of Mechanical Engineering and Materials Science, Duke University.}

\date{\today}

\begin{abstract}
    The interaction of 'walking droplets' and capillary waves in a weakly subcritical Faraday wave experiment has been studied as a hydrodynamic analog to Bohmian quantum mechanics (see "Hydrodynamic Quantum Analogs", J. Bush and A. Oza, Rep. Prog. Physics (2021)). We report here experimental results of walking droplets interacting with supercritical Faraday waves with dimensionless acceleration $\Gamma \approx$ 7.7, where the onset of Faraday instability occurs at $\Gamma_F \approx$ 6.3, in flat bath topography. Our working fluid is silicone  oil with a kinematic viscosity of 20 cst that is placed as a 4.2 mm horizontal liquid layer in an intermediate-aspect-ratio circular bath with a radius to Faraday wavelength ratio of  5.8. We also use different 3D-printed subsurfaces that act as slit structures with local oil depth of 0.7 mm. We confirm expected behavior for walking droplets in the supercritical Faraday regime, such as erratic trajectories, droplets clustering together due to capillary effects, and spontaneous drop creation. We note a special case of walking-droplet behavior when the bath only partially displays Faraday waves. We discuss the influence of the lateral boundaries and slits on droplet trajectory in this chaotic regime and compare the measured trajectories found here to those single and double slit experiments previously studied in the subcritical Faraday regime.
\end{abstract}

\maketitle

\section{\label{sec:level1}Introduction}

Faraday waves are standing wave patterns formed in a bath forced to a sufficient vertical acceleration, hereby noted as the Faraday acceleration $\Gamma_F$, where a bifurcation occurs due to a parametric resonance effect between the bath's natural surface-wave modes and the oscillatory vertical forcing \cite{Miles90}. We will refer to dimensionless vertical acceleration as given by the following equation
\begin{equation}
    \Gamma = \frac{\gamma cos(2\pi ft)}{g}
\end{equation}
where $\gamma = A(2\pi f)^2$, A, f, and t correspond to forcing amplitude, forcing frequency, and time respectively.

\begin{figure}[hbt!]
  \begin{subfigure}{0.23\textwidth}
    \includegraphics[width=\linewidth]{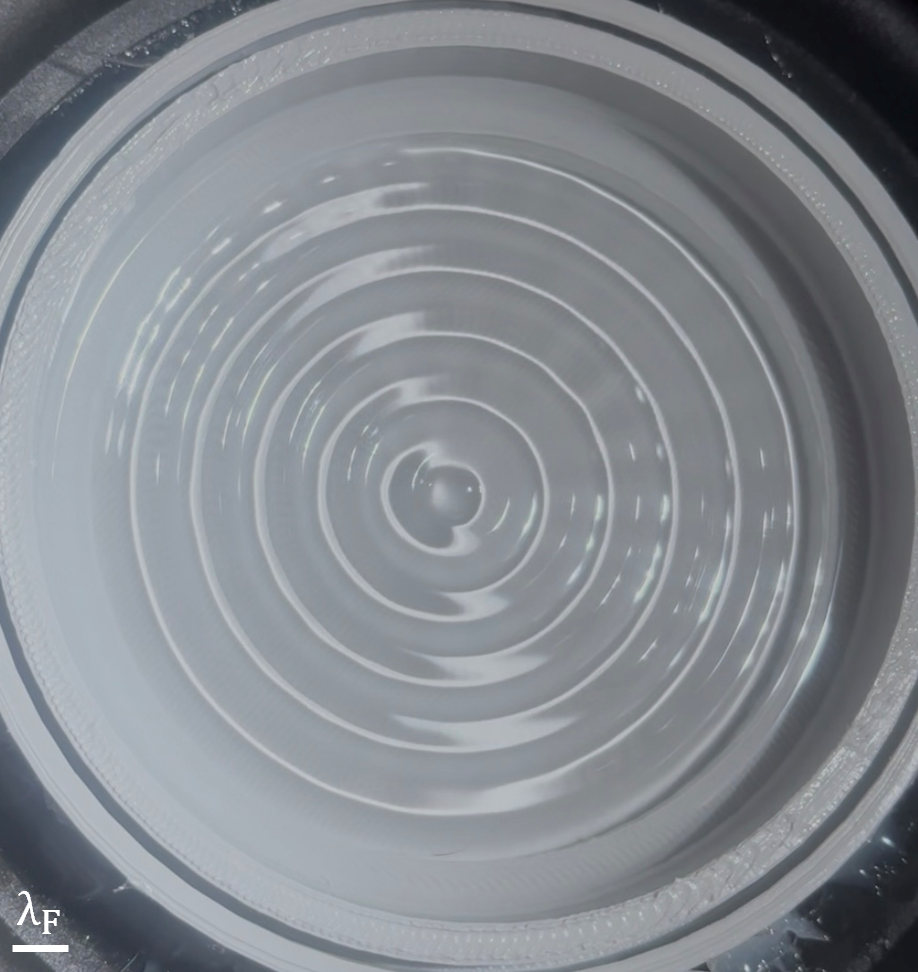}
    \caption{}\label{fig:1a}
  \end{subfigure}%
  \hspace*{\fill}   
  \begin{subfigure}{0.23\textwidth}
    \includegraphics[width=\linewidth]{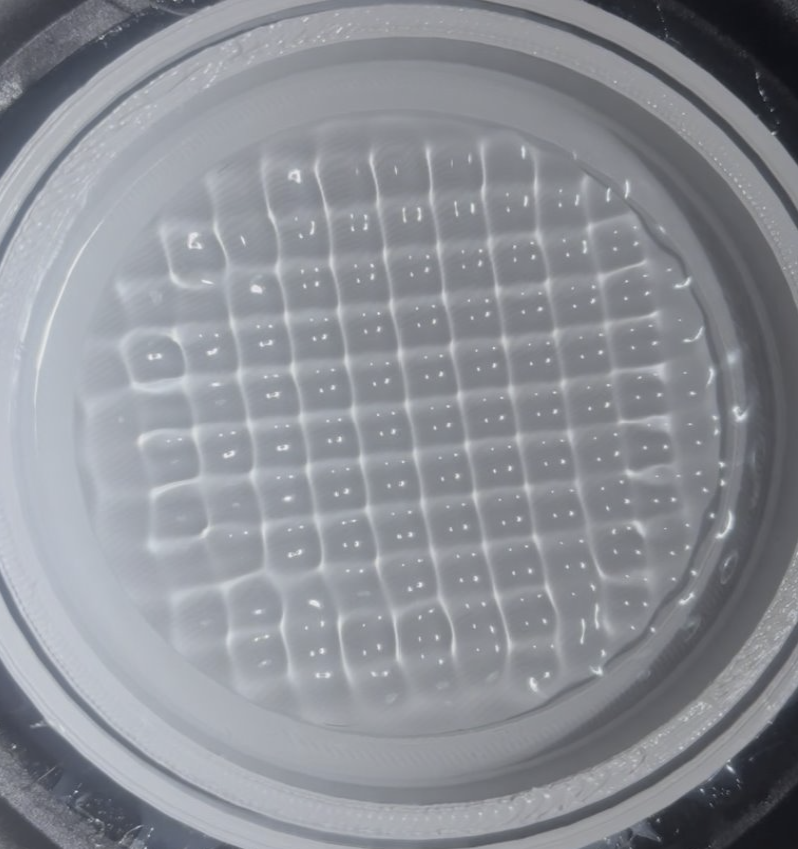}
    \caption{}\label{fig:1b}
  \end{subfigure}%

\caption{Faraday patterns in a vertically forced bath. Here, $\Gamma \approx$ 7.7 and $\lambda_F \approx$ 6.25 mm (a) Radial waves obtained by forcing a bath from a frequency of 0 to 60 Hz at a constant amplitude of 0.6 mm; pattern was transient and quickly fell into square pattern. (b) Square pattern sustained at 60 Hz at a constant amplitude of 0.6 mm.} 
\end{figure}
\hfill \break

Depending on fluid properties and external factors such as the geometry of the bath and forcing amplitude, different patterns form, as seen in figure 1, such as ripples, rectangles, hexagons, or more complex shapes. A rigorous dependence of wave pattern mode on viscosity and forcing frequency is derived from the Navier Stokes equation by Chen $\&$ Vi$\mathrm{\tilde{n}}$als\cite{Chen99} in Ref [2]; figure 2 displays an approximated graph of this dependence with the specific parameters considered in the experiments of this paper. 
\hfill \break

\begin{figure}[hbt!]
    \centering
    \includegraphics[width=1.0\linewidth]{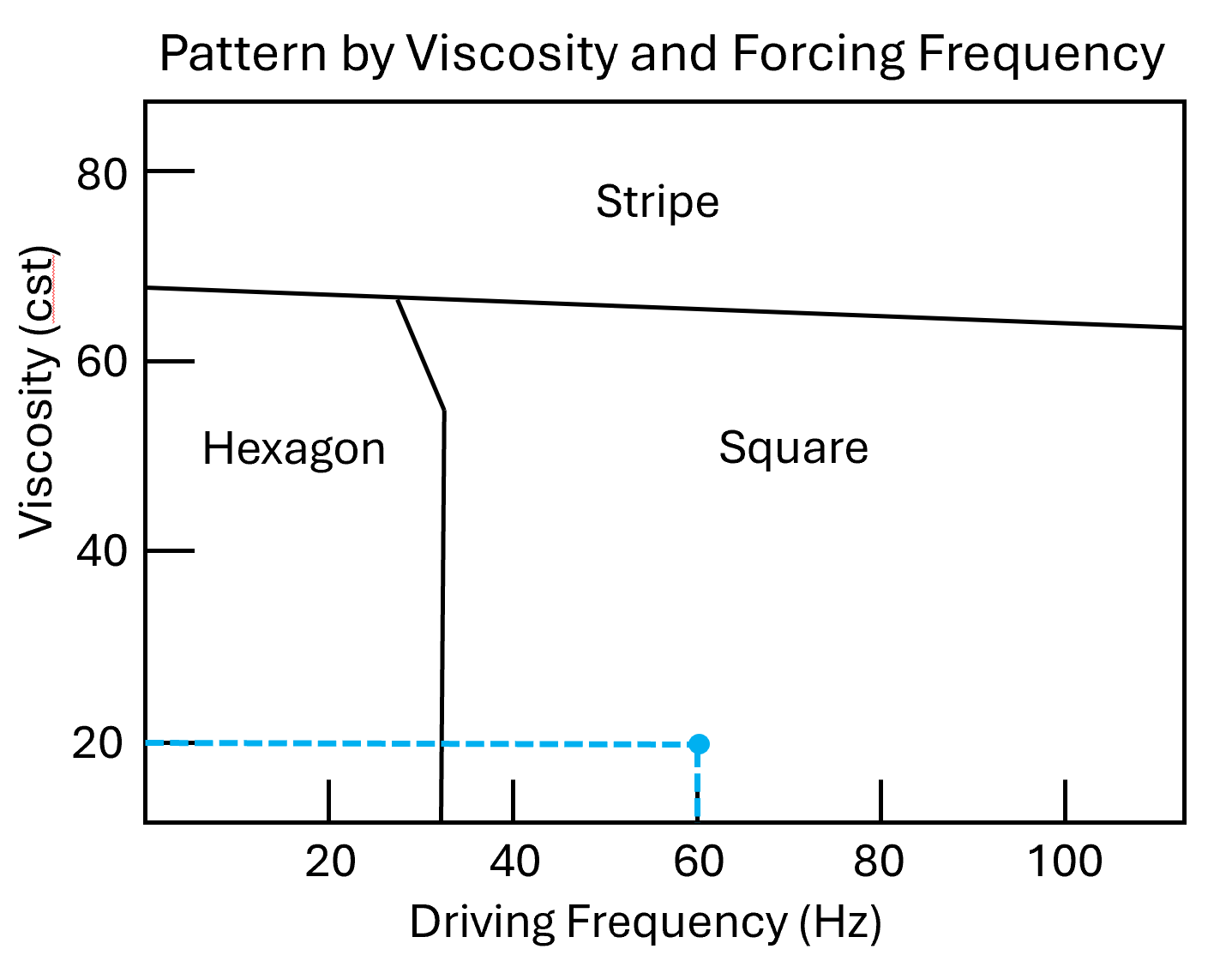}
    \caption{Prediction of patterns based on viscosity and forcing frequency\cite{Chen99} as determined by calculations in Ref [2] for 20cst silicone oil, forcing at 60 hz predicts a square pattern as shown in figure 1b.}
    \label{fig:placeholder}
\end{figure}
When forced well beyond the Faraday acceleration, secondary bifurcations occur, where surface-wave modes compete \cite{Miles90,Cilib84} and the bath can exhibit spatiotemporal chaos (see figure 3).

\begin{figure}[htp!]
    \begin{subfigure}{0.23\textwidth}
    \includegraphics[width=\linewidth]{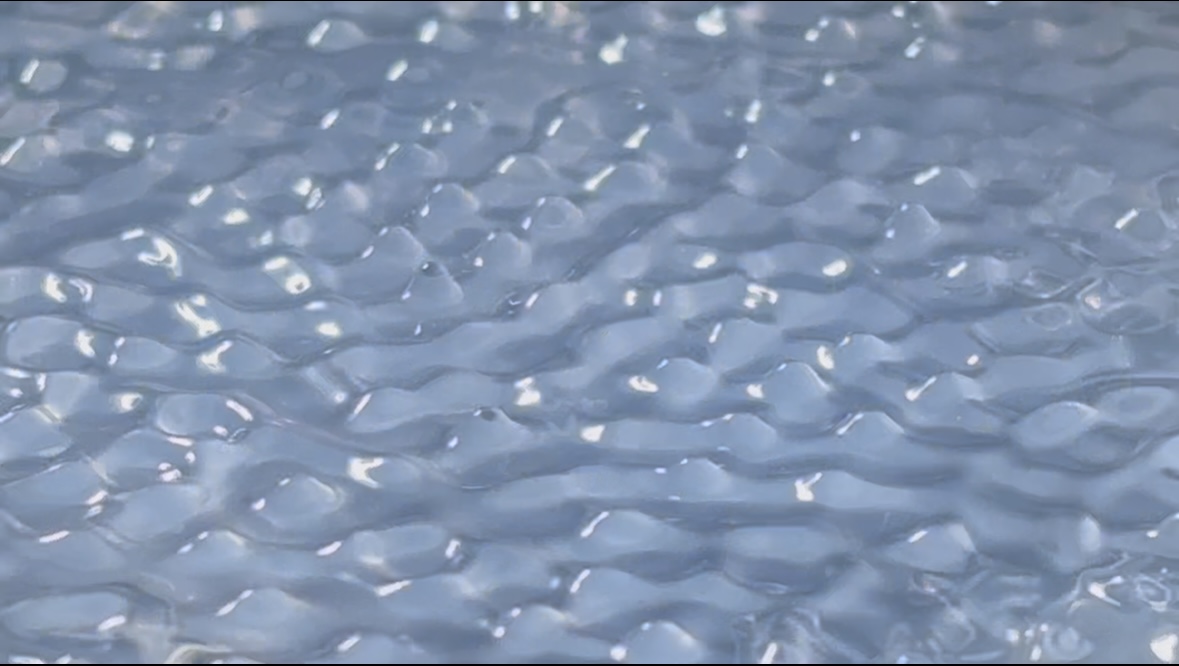}
    \label{fig:1a}
  \end{subfigure}%
  \hspace*{\fill}   
  \begin{subfigure}{0.23\textwidth}
    \includegraphics[width=\linewidth]{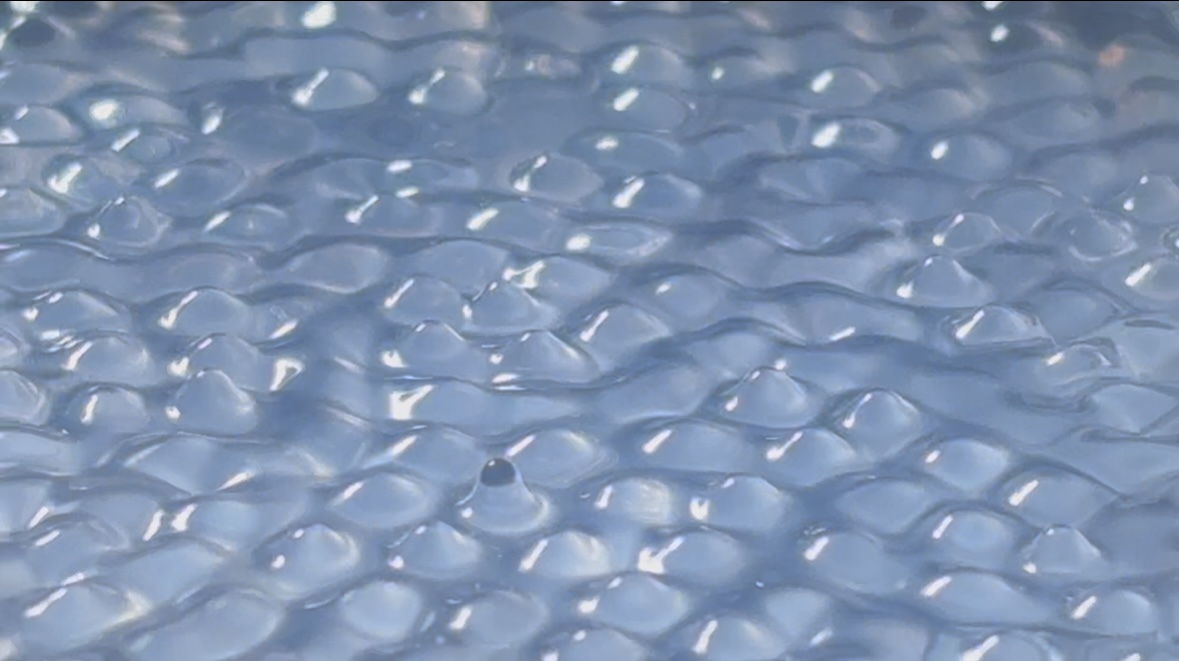}
    \label{fig:1b}
  \end{subfigure}%
  \caption{Faraday waves exhibiting spatiotemporal chaos.}
\end{figure}

In the walking droplet hydrodynamic system, a droplet bounces over vertically forced fluid bath at some acceleration just below the onset of standing Faraday wave patterns, producing its own radial wave field (figure 4). The droplets' bouncing becomes resonant with the Faraday wave mode \cite{Cilib84}, creating a local bifurcation in which the droplet is propelled horizontally by 'walking' over its own wave field.
This system has been studied extensively for behavioral similarities to a plethora of quantum phenomena and qualitative parallels to de Broglie-Bohm pilot-wave quantum theory \cite{CastroBush2023}. Further details of the relationship between walking-droplet dynamics and quantum theory are discussed in Appendix A.
\begin{figure}[htp!]
    \centering
    \includegraphics[width=1.0\linewidth]{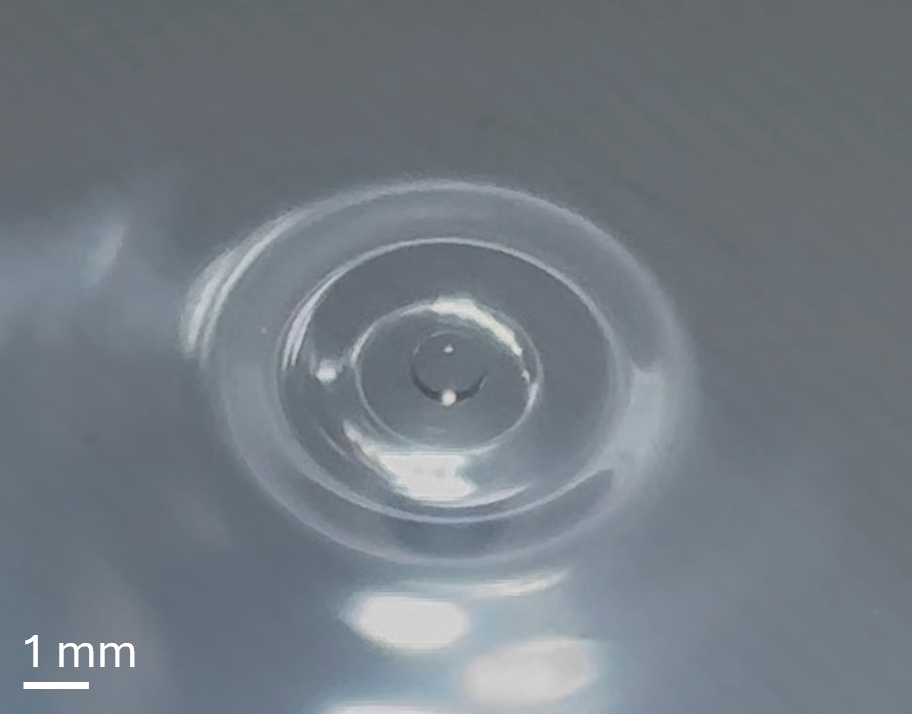}
    \caption{Droplet making its own wave field during vertical bouncing. Here $\Gamma \approx$ 4.0}
    \label{fig:placeholder}
\end{figure}

Notably, while at the heart of wave-particle treatment in mainstream quantum mechanics theory, the diffraction patterns of single- and double-slit quantum experiments have not been successfully re-created in walking-droplet analogs \cite{Bohr15,Pucci18}. To the best of our knowledge, walking-droplet single- and double-slit experiments have been limited to bath regimes of subcritical Faraday acceleration. Majority of walking-droplet experiments are confined to this subcritical regime, though the experiments of Tambasco et al. in Ref [6] provide detailed results of walking-droplet behavior in baths forced above the Faraday threshold \cite{Tamb18}.

 In this paper, we first document and explore walking-droplet behavior in the supercritical Faraday acceleration regime and then present novel single- and double-slit experiments for a bath forced above the Faraday wave pattern threshold. The primary goals of this paper are to demonstrate the rich dynamics of supercritical walking-droplet experiments, and to detail an inexpensive experimental setup along with simple data analysis methods.
\section{Methods}
\subsection{Apparatus}
We constructed an apparatus adapted from Harris et al.\cite{Harris17} in Ref [8] consisting of a container of 20 cSt silicone oil adhered to a subwoofer receiving an amplified sinusoidal signal from an amplifier; a lab camera was used for data collection with a white lamp nearby the subwoofer for improved visualization (figure 5). In our apparatus, the sinusoidal signal was produced connecting the amplifier to a laptop via audio cable, and producing a sine wave from an online tone generator (we used https://www.szynalski.com/tone-generator/), where frequency is toggled incrimentally and vertical amplitude corresponded to laptop output volume.

\begin{figure}[htp!]
    \centering
    \includegraphics[width=1.0\linewidth]{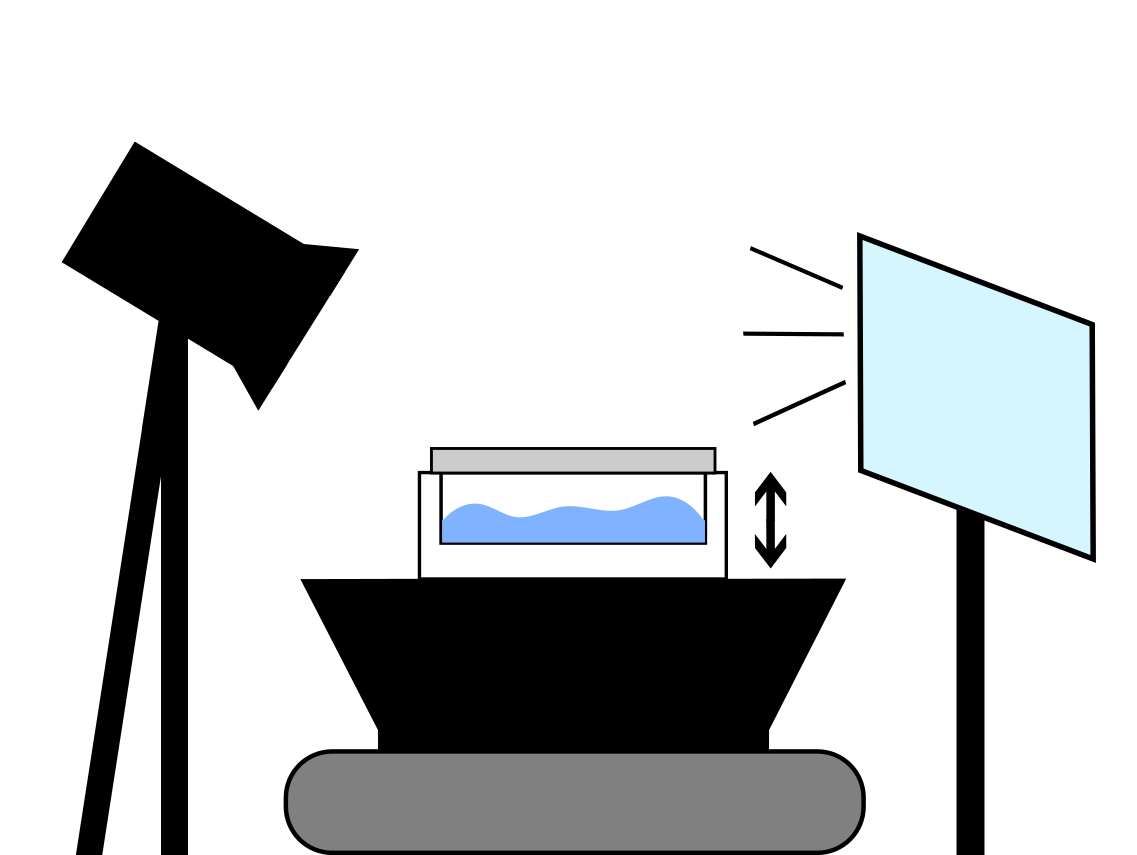}
    \caption{A bath of silicone oil experiences forced vertical oscillation via coupling to a subwoofer. The subwoofer provides the vertical oscillation in response to an amplified sinusoidal signal from the amplifier.}
    \label{fig:placeholder}
\end{figure}

We note that in this inexpensive setup, the subwoofer imperfectly translates the sinusoidal source into vertical motion. This imperfection is apparent in lower frequencies (1-10 Hz) and is demonstrated in figure 6. At higher frequencies, the disparity is less apparent due to the smaller time spent at vertical extrema. Though the vertical forcing is not perfectly sinusoidal, our experimental setup produces the predicted rectangular patterns, as suggested by the agreement between figure 1b and figure 2.

\begin{figure}[htp!]
    \centering
    \includegraphics[width=1.0\linewidth]{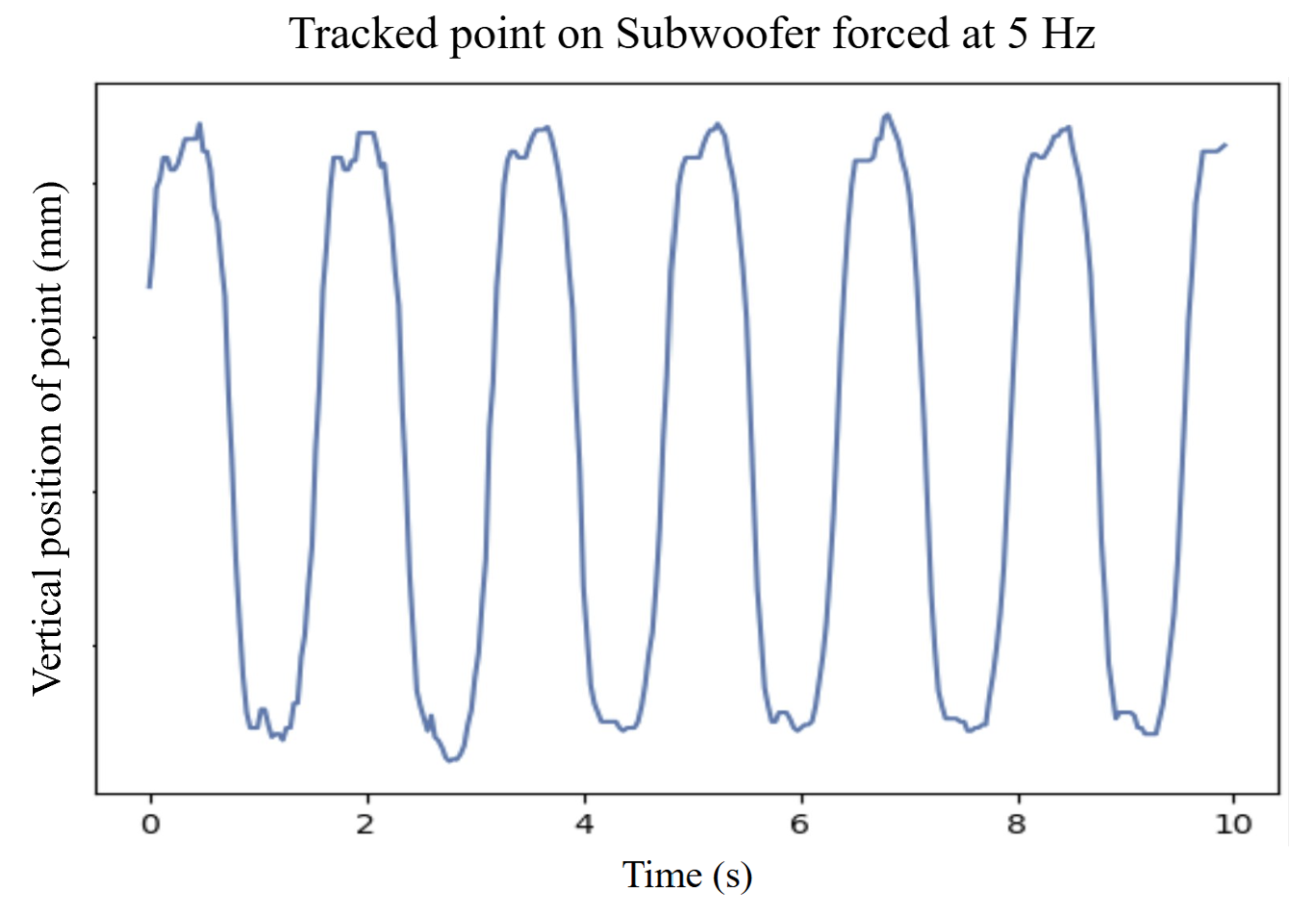}
    \caption{By tracking a point on the subwoofer forced at low frequency 5 Hz, it is clear the subwoofer provides an imperfect sinusoidal forcing that may introduce additional nonlinear effects.}
    \label{fig:placeholder}
\end{figure}

We 3D printed various fluid bath containers, differing in diameter and inner base topography depending on the corresponding experiment. All containers were printed with a concentric, shallow, millimeter-wide indentation on the rim of the bath to secure a Petri dish lid, which prevents the influence of air currents on droplet trajectories. In Figure 7, we note a concentric layer of 3 mm shallower depth than the central bath, which we hereby refer to as a 'boundary layer', intended to deter pattern formation at the edge of the container and negate the dynamical influence of the container wall.

\begin{figure}[htp!]
    \centering
    \includegraphics[width=1.0\linewidth]
{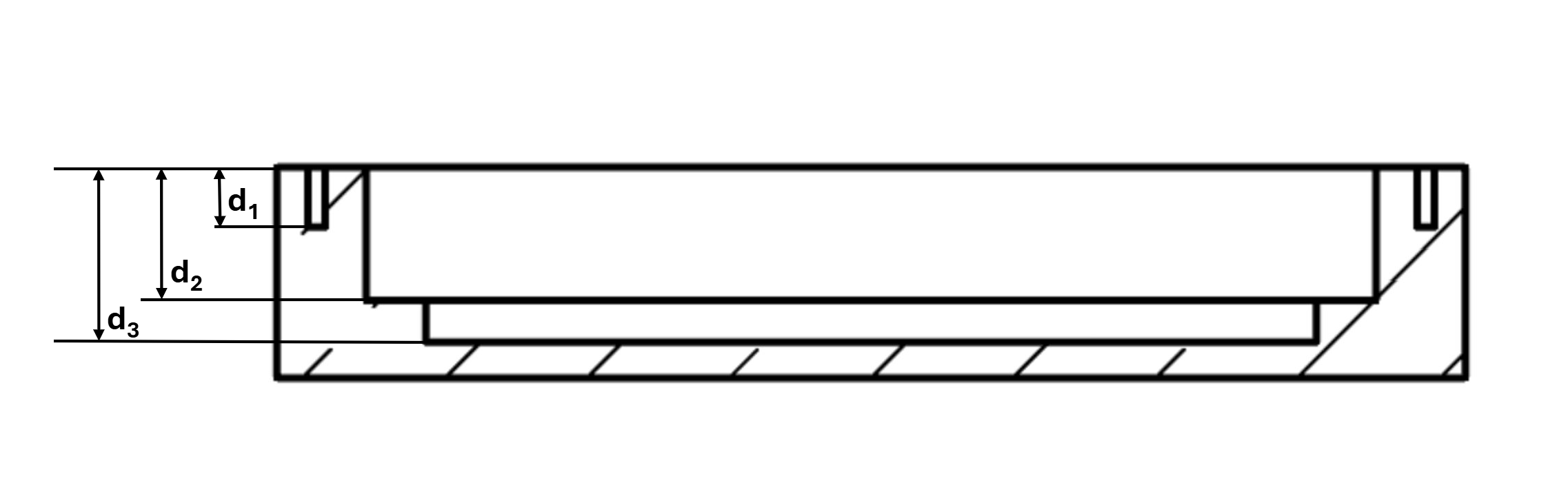}
    \label{fig:placeholder}
  \caption{Section view of circular bath container with significant depths labeled as $d_1 - d_3$. The indentation of $d_1$ = 5.0 mm is a millimeter-wide slit to allow the insertion of a translucent petri dish lid onto the container, preventing air currents from affecting droplet trajectory. The indentation $d_2$ = 11.5 mm corresponds with the boundary layer along the outer rim of the bath bottom. $d_3$ = 14.5 mm is the greatest bath depth where majority of the fluid lay and where nonlinear wave patterns will first develop.}
\end{figure}

In experiments without slits, the inner bath base is radially symmetric, with all sections about the center axis being identical to figure 7. We adapted the bath topography for experiments with slits from Anderson et al.\cite{Bohr15} in Ref[5]. As shown in figure 8, slits are constructed by extending the boundary layer in as a 'wall' in the approximate center of the bath. Slits are 5 mm wide indentations at the depth of the larger bath. In the single-slit case, one slit is centered, where in the double-slit case, two slits are centered with a 5 mm boundary layer depth section between the slits. For both single- and double-slit containers, the upper region of the bath includes an extended boundary layer with a triangular indentation intended to promote droplets' motion toward the slit(s).

\begin{figure}[htp!]
    \begin{subfigure}{0.23\textwidth}
    \includegraphics[width=\linewidth]{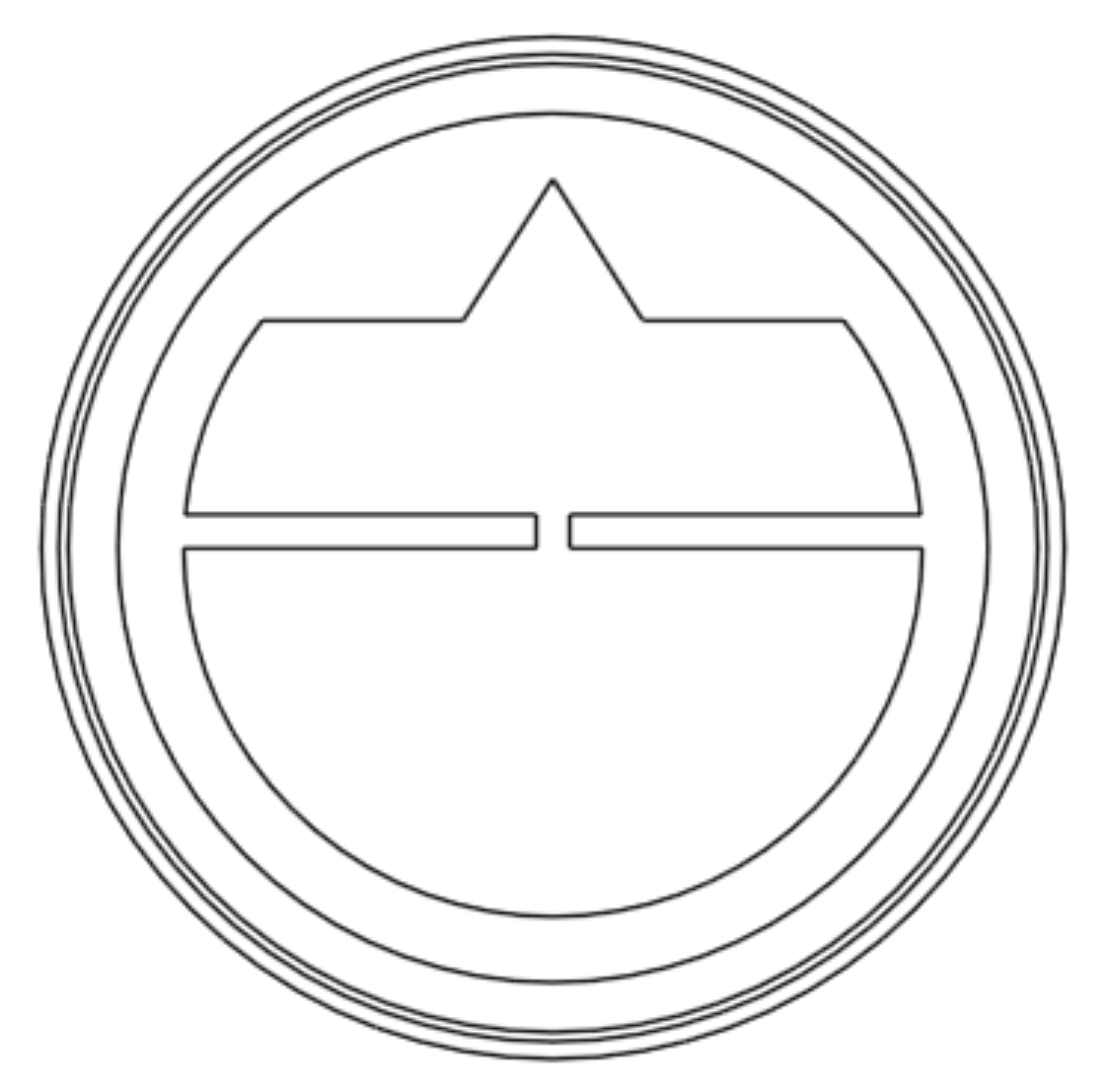}
    \label{fig:1a}\caption{}
  \end{subfigure}%
  \hspace*{\fill}   
  \begin{subfigure}{0.23\textwidth}
    \includegraphics[width=\linewidth]{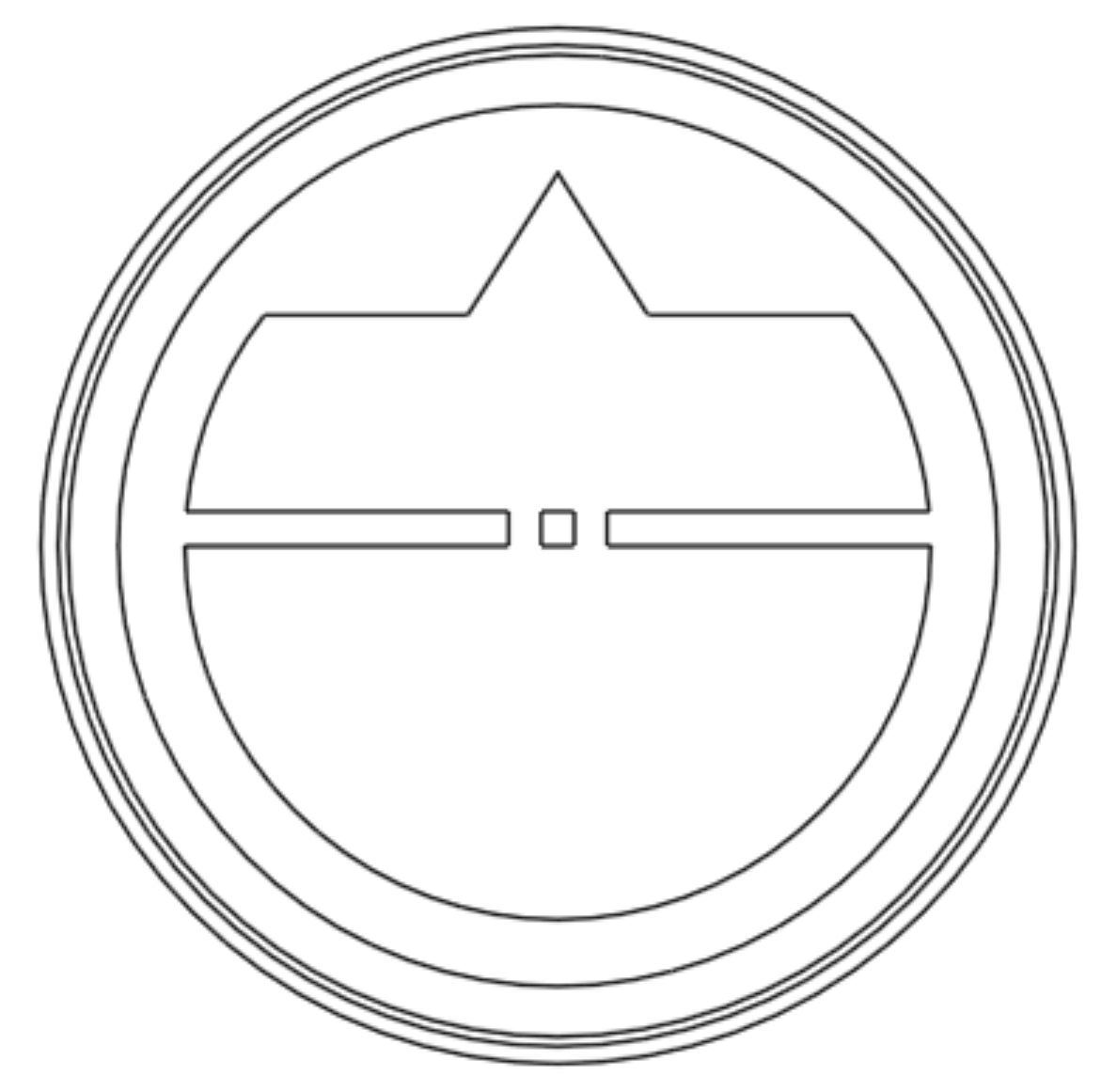}
    \label{fig:1b}\caption{}
  \end{subfigure}%
  \caption{Top view of single(a) and double(b) slit experiment bath container topography. 'Walls' around slits are of an equal depth to the boundary layers, at 11.5 mm, while 'slits' are at an equal depth to the greater bath, at 14.5 mm. The single slit is a width of 5 mm, and the double slits are each 5 mm wide with a separation of 5 mm. The triangular portion at the top of both (a) and (b) is intended to guide the droplets toward the slits when one is placed during vertical forcing.}
\end{figure}
\subsection{Non-slit Experiments}
In initial experiments, baths with outer diameter of 155 mm, inner rim diameter of 140 mm, and boundary layer edge diameter of 112.5 mm, held silicone oil at a depth of 4.2 $\pm$ 0.2 mm in the center region and 1.2 $\pm$ 0.2 mm in the boundary layer region. The bath was coupled to a subwoofer providing vertical forcing at varying amplitude. For a strobe effect to allow slowed visualization of Faraday waves, we set the lab camera to record video data at 60 Hz and used an offset subwoofer forcing frequency of 59 Hz.
We began experiments by forcing the subwoofer at a low amplitude where droplets displayed vertical bouncing only. Droplets were placed on the bath via tapping with a micro-pipette (we note that any pointed, needle-like object would be sufficient for creating drops with this method). After placing the horizontally-stationary droplets, we secured the Petri dish lid onto the outer indentation. We then slowly increased amplitude until Faraday wave patterns appeared.
In these 'larger' bath experiments with outer diameter of 155 mm, wave patterns tended to favor forming on one side, while the other side had a 'flat' wave topography. This asymmetry is discussed further in the 'Results' section. We do not report a $\Gamma_F$ for these results, as different regions of the bath appear to be at different effective dimensionless accelerations, making it difficult to quantify the Faraday acceleration for the full bath.
Due to this asymmetric Faraday wave topography, we decreased the bath dimensions, where all other experiments were done with bath outer diameter of 100 mm, inner rim diameter of 85 mm, and boundary layer edge of 75 mm. We repeated the experimental procedure above for a bath with no slits and these smaller dimensions. Here, Faraday patterns filled all regions of the bath evenly at the same time. For a bath of this sized, filled to depth 4.2 mm $\pm$ 0.2 mm with boundary depth 1.2 mm $\pm$ 0.2 mm, we found an onset of Faraday waves occurred at a vertical forcing amplitude of 0.45 mm for a calculated $\Gamma_F \approx$ 6.3. 
We slowly increased forcing amplitude and recorded walking-droplet behavior over rectangular Faraday wave patterns at $\Gamma \approx$ 7.7 and over chaotic Faraday waves at $\Gamma \approx$ 9.8. We continued increasing amplitude until spontaneous droplet formation occurred at $\Gamma \approx$ 11.2.

\subsection{Slit Experiments}
In experiments with slits, bath containers were of the smaller dimensions described in the previous subsection. In each experimental run, we force the bath at an amplitude $\approx$ 0.34 mm with $\Gamma \approx$ 4.8 where droplets bounce in the vertical direction only. We use a micropipette to place a droplet in the upper region before the slit(s), as indicated in figure 9. 

\begin{figure}[htp!]
    \centering
    \includegraphics[width=1.0\linewidth]{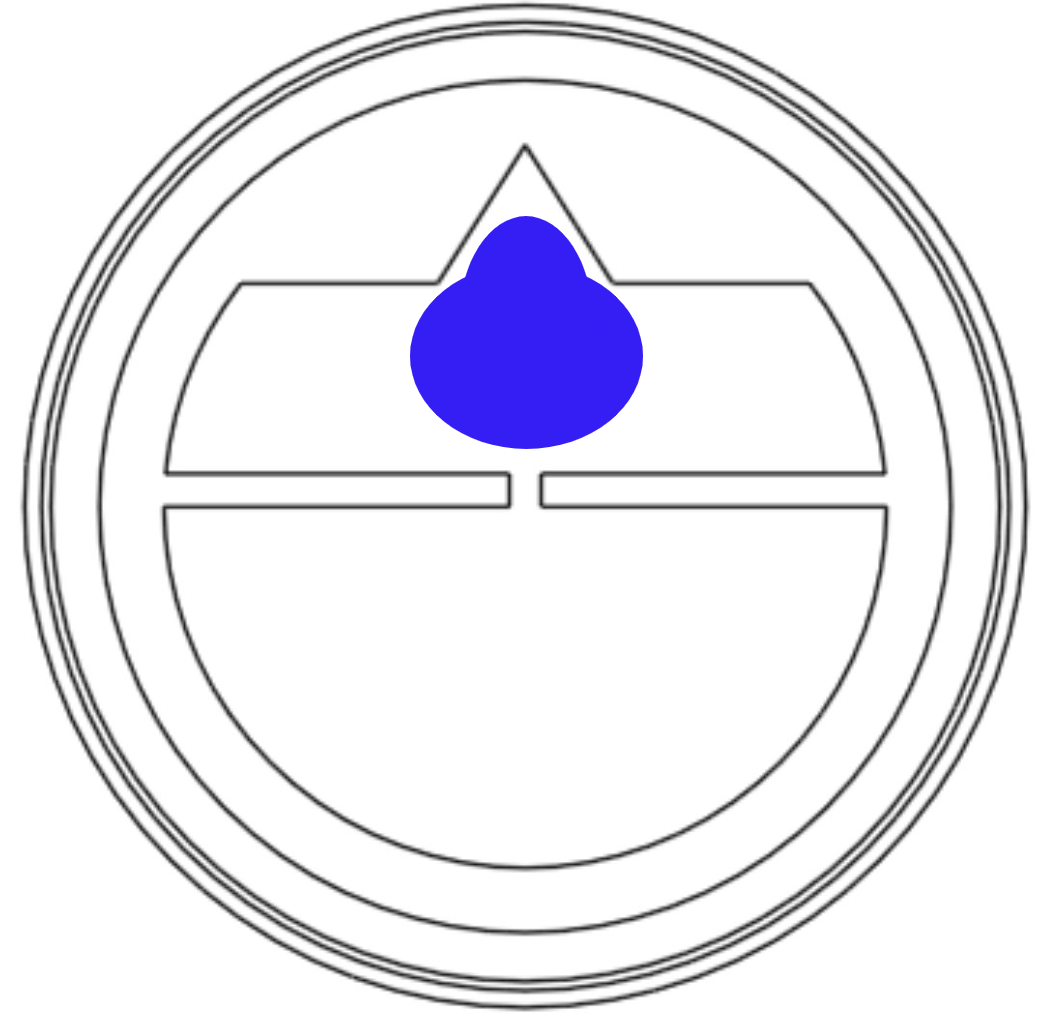}
    \caption{Region where droplets are placed in single- and double-slit experiments.}
    \label{fig:placeholder}
\end{figure}

We then secure the Petri dish lid and increase the forcing amplitude to $\Gamma \approx$ 7.7 where rectangular Faraday waves appear in the deepest regions of the bath.

\subsection{Data Analysis Methods}

Trajectories were plotted with the Vernier Video Physics app, in which video data is uploaded to the app, and the drop position is selected frame-by-frame. We input the appropriate scale on the app to generate a trajectory graph. The trajectory was then scaled and formatted over images of the appropriate bath, often where the droplet was in its last position. We note the pedagogical value such an app has given its low cost and simplicity.
Other methods of droplet tracking were attempted using open source computer vision software, and we found them to be insufficient in tracking droplets in our video data. We discuss these attempts further in Appendix B.
\section{Results}
\FloatBarrier
We present observations of walking-droplet behavior in the no-slit experiments, first where the entire bath displays Faraday waves, and second where certain regions of the bath display Faraday waves before other regions. We then present observations and trajectories of single- and double-slit experiments in the supercritical Faraday acceleration regime. For all of the 'small' baths (outer diameter of 100 mm), $\lambda_F \approx$ 6.25 mm, and for the 'large' baths (outer diameter 155 mm) $\lambda_F \approx$ 7.03; this was measured by the number of Faraday waves per greatest-depth bath diameter.

\subsection{Droplets in Radially Symmetric Faraday Wave Topography}

\begin{figure}
\includegraphics[width=1.0\linewidth]{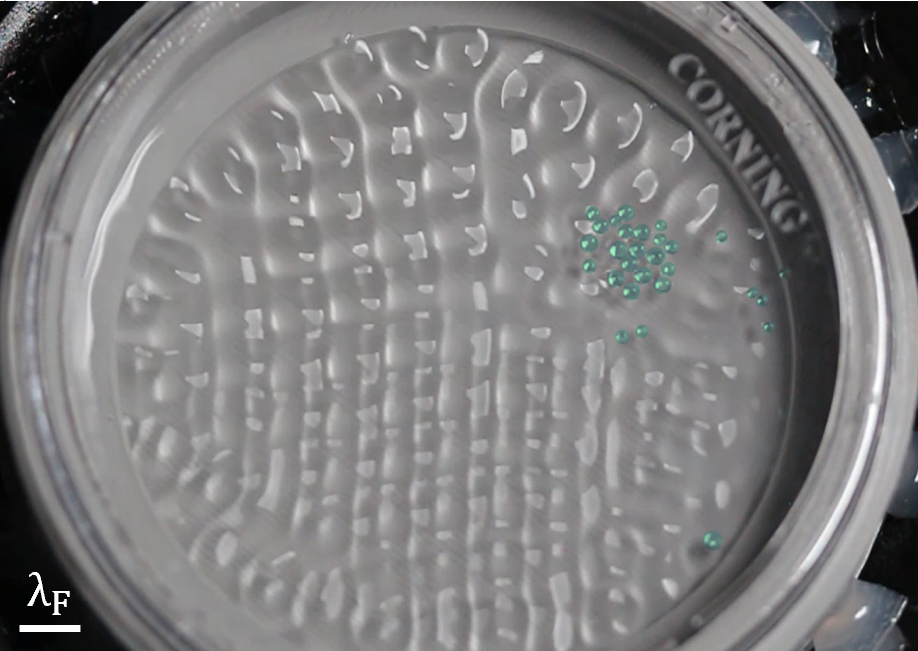}
\caption{\label{fig:epsart} Highlighted drop cluster in bath exhibiting nonlinear patterns at $\Gamma \approx$ 7.7.}
\end{figure}

In baths forced at $\Gamma \approx$ 7.7, we find that walking droplets migrate towards each other, and tend to form clusters due to the lattice-like minima and maxima of the rectangular Faraday pattern. At this acceleration, droplets tend to stay horizontally fixed in this cluster, eventually coalescing with the bath (fig 10).
At $\Gamma >$ 7.7, we find that droplets initially in clusters will begin to break out, occasionally forming new clusters (fig 11). Here, as the Faraday waves are transitioning toward spatiotemporal chaos, droplets will occasionally bounce resonantly with a Faraday mode that is different from the dominating square pattern, allowing it to leave the cluster. As the square pattern is still the dominant mode, clusters are still forming. At $\Gamma >> \Gamma_F$ (for our apparatus, around $\Gamma \approx$ 9.8) when Faraday waves fall out of the square lattice and into chaos, clusters are no longer able to form.

\begin{figure}[htp!]
    \begin{subfigure}{0.23\textwidth}
    \includegraphics[width=\linewidth]{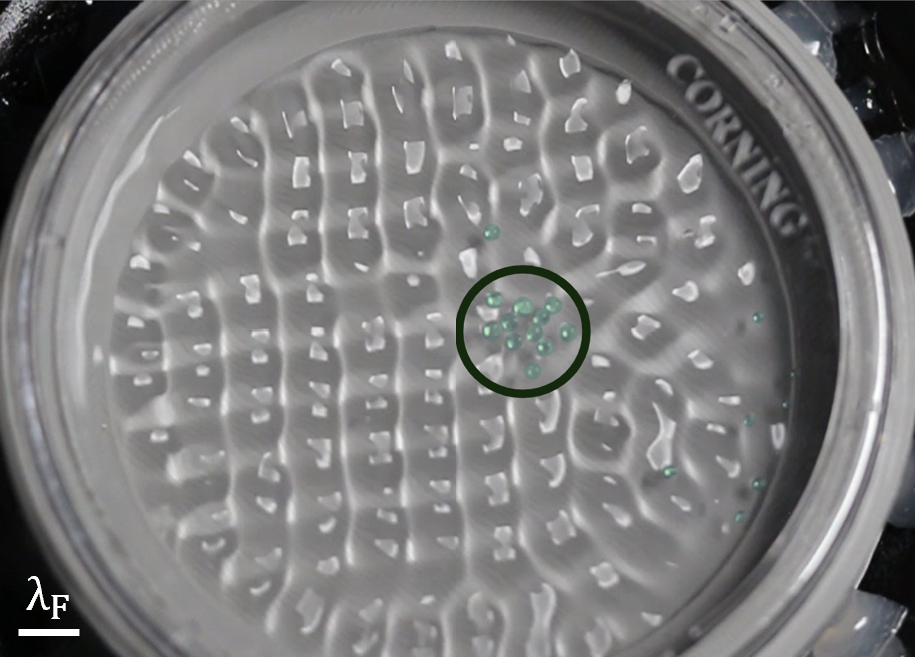}
    \label{fig:1a}\caption{}
  \end{subfigure}%
  \hspace*{\fill}   
  \begin{subfigure}{0.23\textwidth}
    \includegraphics[width=\linewidth]{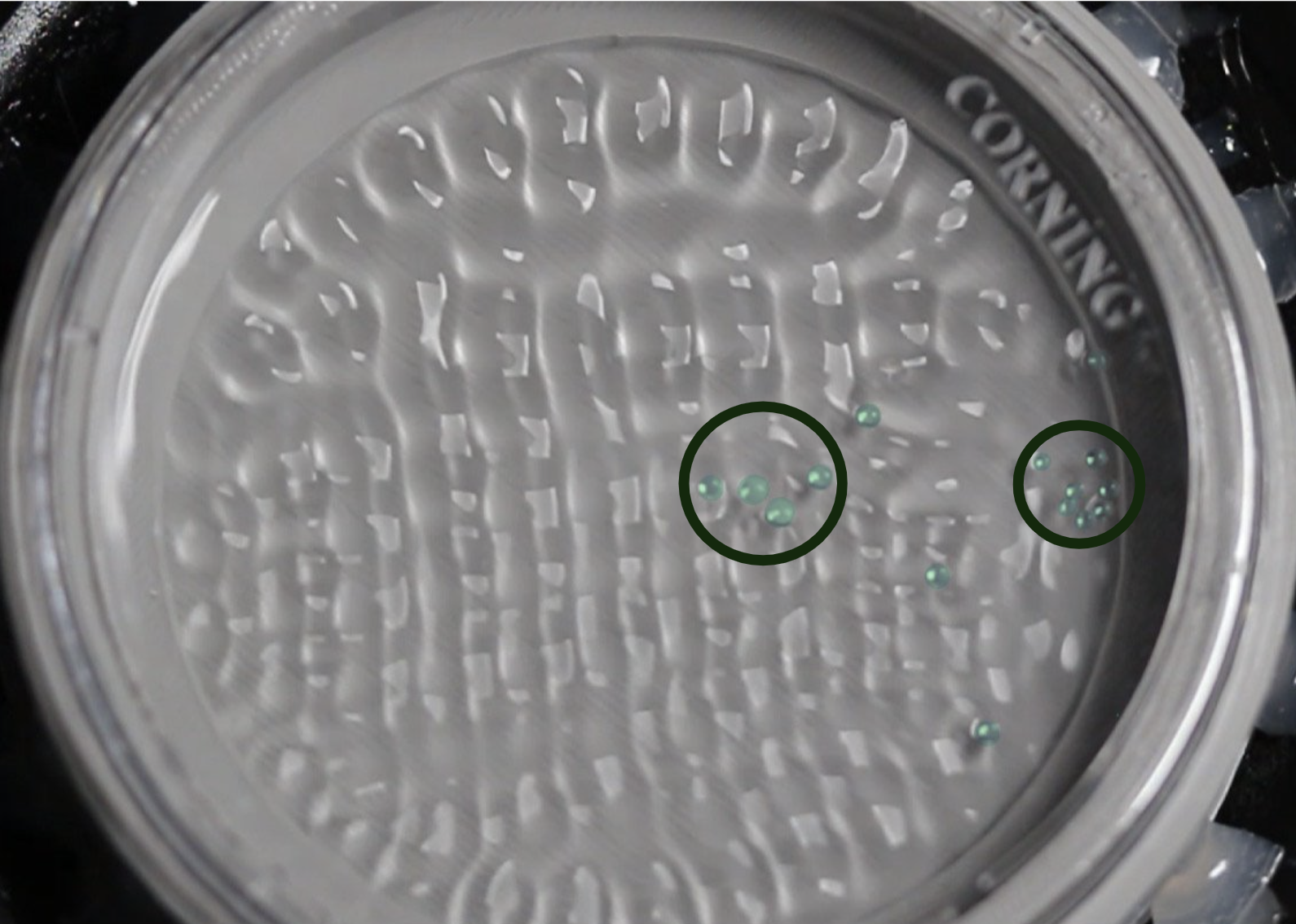}
    \label{fig:1b}\caption{}
  \end{subfigure}%
  \caption{Beyond the onset of nonlinear patterns ($\Gamma \approx$ 9.5), droplets break out of clusters and form new clusters. (a) and its subsequent frame (b) show droplets leaving a cluster and reforming a new cluster on the far right side of the fluid bath.}
\end{figure}

At sufficiently high Faraday acceleration, in our experiments around $\Gamma \approx$ 11.2 (fig 12), the bath began to spontaneously form droplets (fig 13). In this regime, vertical forcing is large enough that gravity and surface tension are no longer able to prevent the peak of a Faraday wave from spontaneously breaking off. 

\begin{figure}[htp!]
    \centering
    \includegraphics[width=1.0\linewidth]{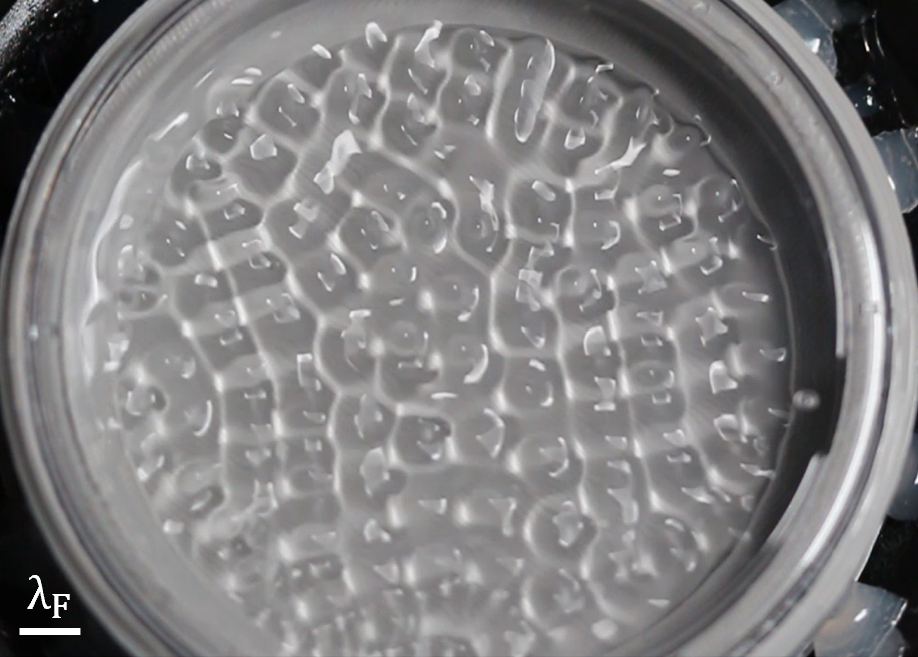}
    \caption{Nonlinear wave patterns well beyond the onset of spatiotemporal chaos ($\Gamma \approx$ 11.2) at the cusp of the spontaneous drop-forming regime.}
    \label{fig:placeholder}
\end{figure}

\begin{figure}[htp!]
    \begin{subfigure}{0.23\textwidth}
    \includegraphics[width=\linewidth]{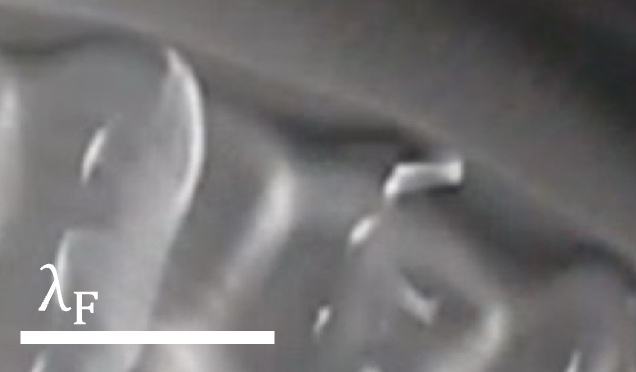}
    \label{fig:1c}\caption{}
    \end{subfigure}%
    \hspace*{\fill}
    \begin{subfigure}{0.23\textwidth}
    \includegraphics[width=\linewidth]{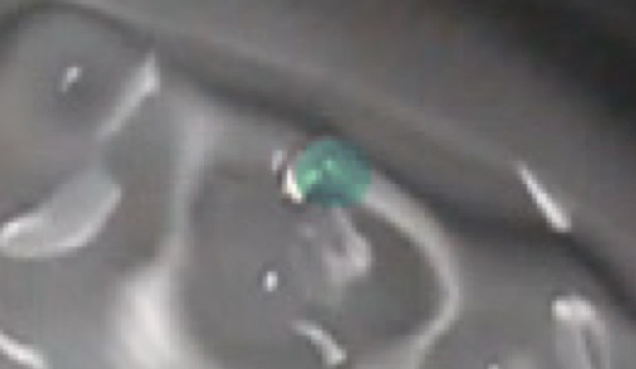}
    \label{fig:1d}\caption{}
    \end{subfigure}%
    \hspace*{\fill}
    \caption{A close up of the top-right corner of the circular bath forced well beyond the critical Faraday acceleration ($\Gamma \approx$ 11.2) from the previous figure (a). In this regime, drops spontaneous form and coalesce with the greater bath (b).}
\end{figure}

Beyond $\Gamma_F$ droplets meander, walking with directional changes over lengths greater than the Faraday length, and move erratically, with directional changes within a few bounces, on the order of the Faraday wavelength (fig 14).

\begin{figure}[htp!]
    \centering
    \includegraphics[width=1.0\linewidth]{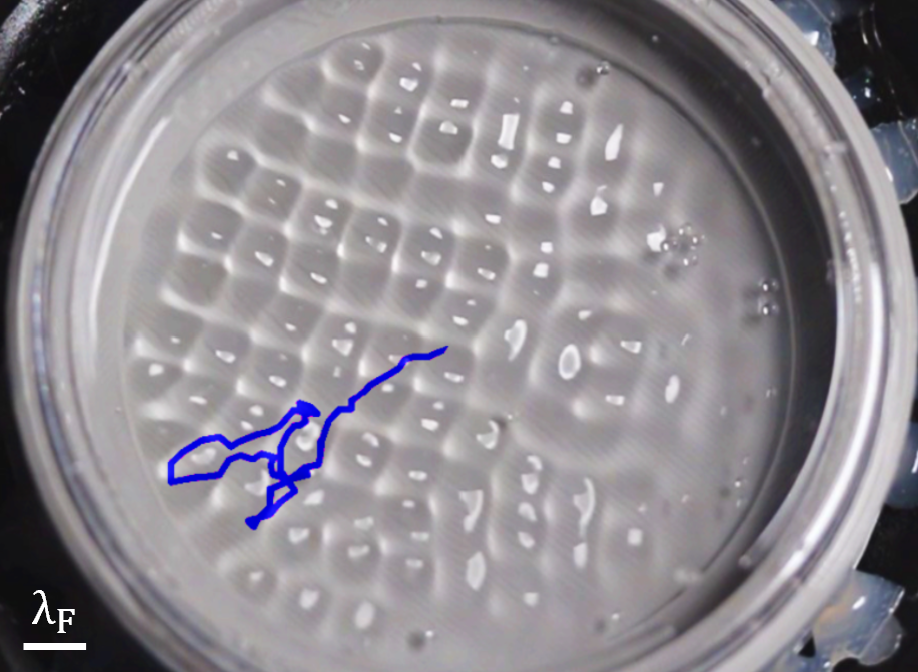}
    \caption{Erratic trajectory of a walking droplet bouncing over a fluid bath exhibiting patterns of spatiotemporal chaos. Trajectory is formatted over final droplet position. ($\Gamma \approx$ 9.5)}
    \label{fig:placeholder}
\end{figure}

\FloatBarrier
\subsection{Droplets in Asymmetric Nonlinear Wave Topography}

In larger fluid baths of 112.5 mm inner diameter, Faraday wave patterns appeared in certain regions of the bath while other regions remain without patterns. Forced well beyond the onset of these patterns in a certain regime, the region where patterns initially rise spontaneously formed droplets, while a region on the opposite end of the bath remained without wave patterns (fig 15). The reason for the asymmetric onset of Faraday wave patterns was not thoroughly investigated in this project, though we suspect asymmetry may be due to boundary interactions between the subwoofer and adhesed container.

\begin{figure}[htp!]
    \centering
    \includegraphics[width=1.0\linewidth]{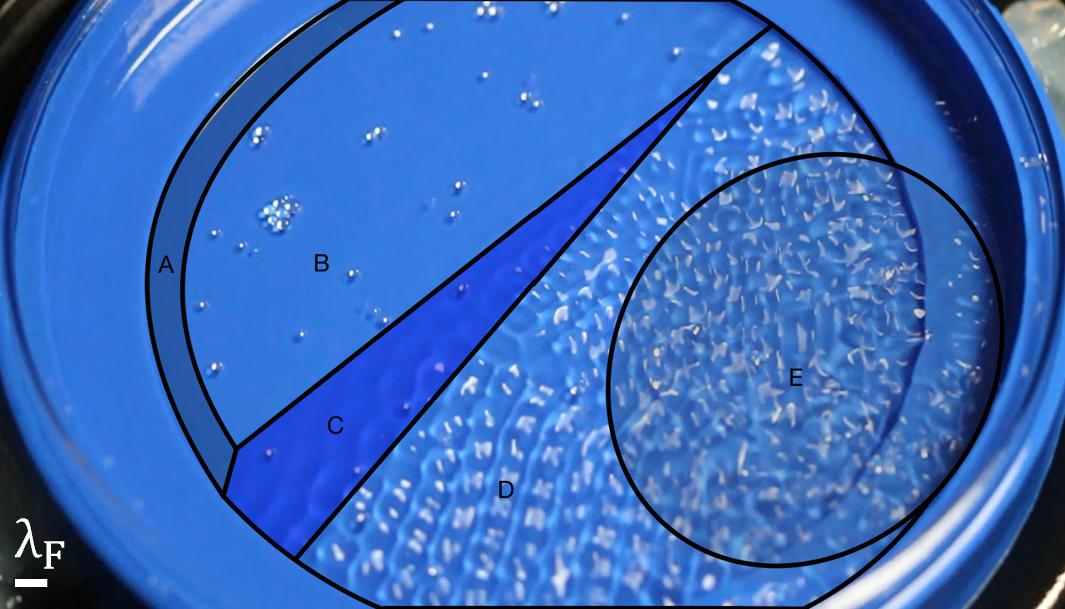}
    \caption{Regions of drop behavior in a bath with asymmetrical nonlinear pattern topography, organized by behavior due to Faraday acceleration. Across Region B, moving towards Region A, the drops experience increasingly less Faraday acceleration, causing drops to eventually remain horizontally stationary, bouncing only in the vertical direction. This causes Region A to be void of droplets, at the lowest Faraday acceleration in the bath. Region C is an interfacial region between visible nonlinear wave topography and 'flat' fluid topography. Region D shows nonlinear square wave patterns, while Region E had reached a Faraday acceleration sufficient for spatiotemporal chaos and spontaneous drop formation.}
    \label{fig:placeholder}
\end{figure}

In experiments with asymmetric nonlinear wave pattern topography, we observed that droplets tend to walk away from regions of visible wave patterns and toward regions without patterns. In figure 16, a droplet has spontaneously formed in a region of chaotic wave patterns, and walks toward a calmer region, where they move forward horizontally until below the acceleration needed for horizontal walking.

\begin{figure}[htp!]
    \centering
    \includegraphics[width=1.0\linewidth]{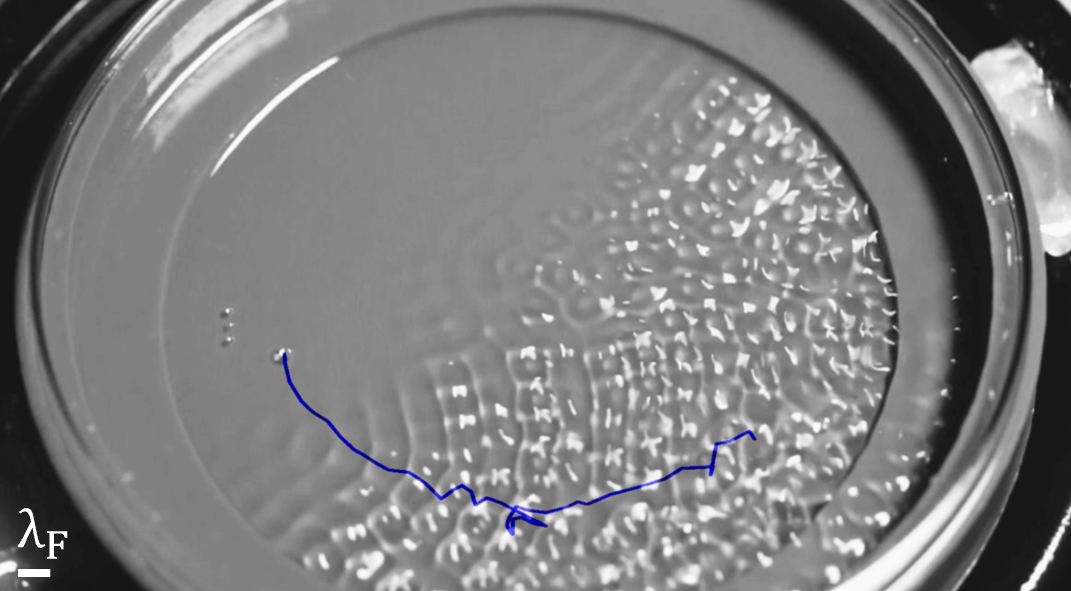}
    \caption{In a fluid bath with asymmetrical nonlinear wave topography, spontaneously formed droplets move from region of spatiotemporal chaos to region without nonlinear patterns. Trajectory begins in region of chaotic waves and is formatted over the final drop position.}
    \label{fig:placeholder}
\end{figure}

We found that in a case where the entire bath displayed Faraday patterns, the region of initial pattern onset began to spontaneously produce droplets. The droplets again preferred to move away from this region (fig 17). 

\begin{figure}[htp!]
    \centering
    \includegraphics[width=1.0\linewidth]{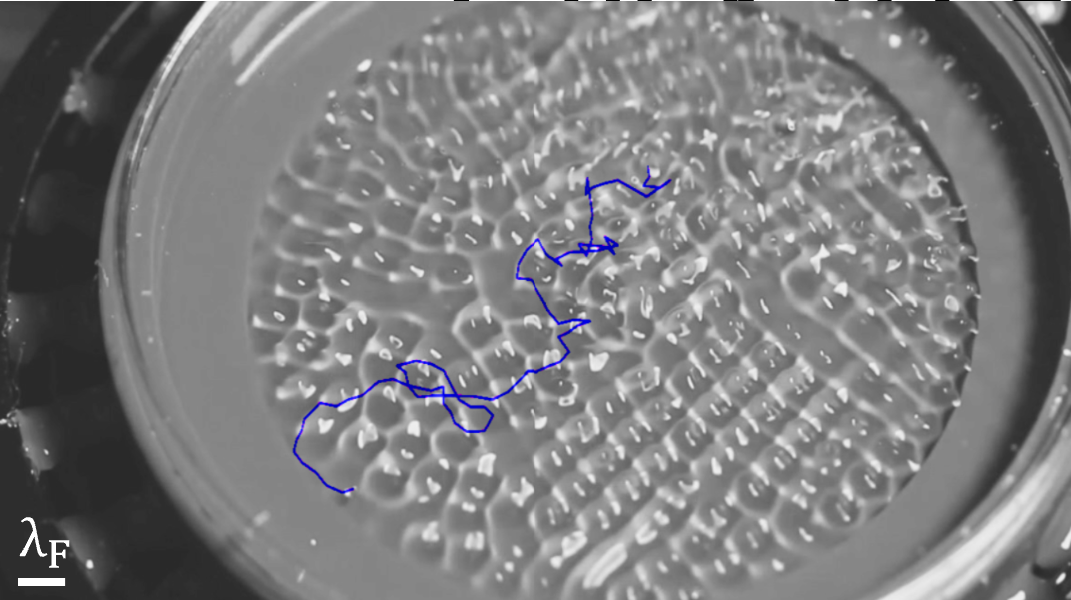}
    \caption{When entire bath displays nonlinear pattern with asymmetrical topography, a spontaneously formed droplet walks to the bath region with less-chaotic topography. Trajectory of spontaneously formed droplet walking over asymmetric topography of bath displaying nonlinear patterns (formatted over final droplet position).}
    \label{fig:placeholder}
\end{figure}

Before continuing with smaller diameter containers, we conducted droplet experiments with a large (112.5 mm inner diameter) container with a single-slit configuration. In this earlier iteration of slit configuration, the boundary layer was of 3 mm greater depth in comparison with the larger fluid bath and with the centered slit (inverse with later iterations of shallow boundaries and deep slits). We found that again one region of the bath displayed Faraday patterns first and spontaneously formed droplets. Additionally, the boundaries and walls leading to the slit saw the onset of Faraday waves. Walking droplets were averse to staying within these boundaries and meandered over the unpatterned bath regions, going through the slit, and continuing to walk until reaching a region where the droplet only moved vertically (fig 18).

\begin{figure}[htp!]
    \centering
    \includegraphics[width=1.0\linewidth]{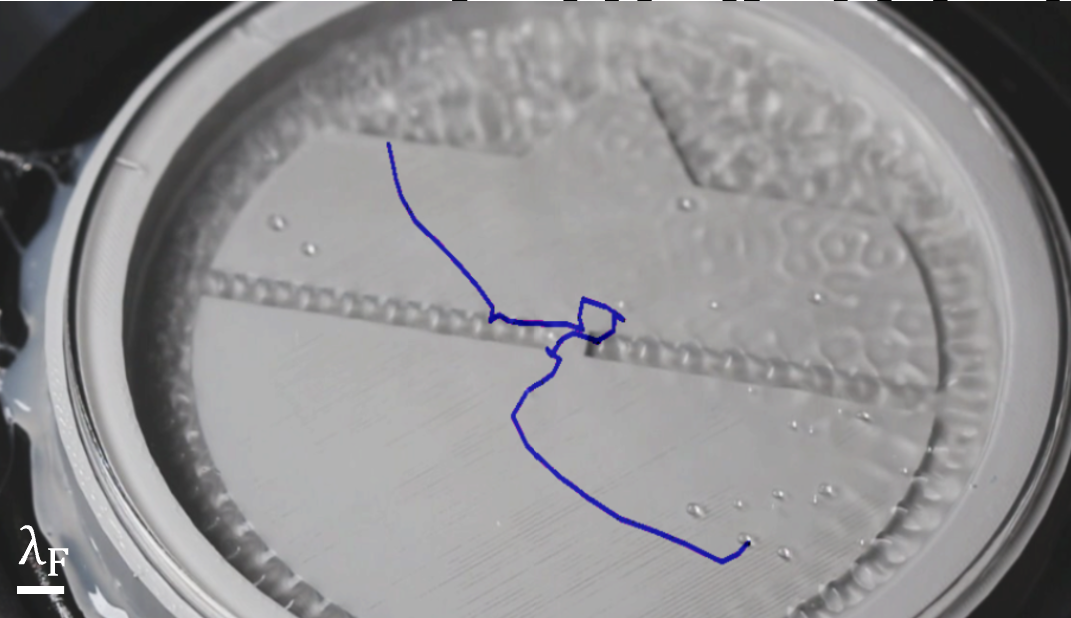}
    \caption{A bath with single-slit experiment indentation displayed asymmetric nonlinear wave topography. Drops spontaneously formed in the top left region, traversed through the slit and continued walking horizontally until reaching a region of solely vertical bouncing on the far right.}
    \label{fig:placeholder}
\end{figure}
\FloatBarrier
\subsection{Droplets in Supercritical Single-Slit Experiments}
In single-slit experiments for the 85 mm inner diameter bath, we observed that Faraday patterns appeared in deeper regions, where the shallow boundaries and 'walls' leading to the central slit did not have patterns. At $\Gamma \approx$ 7.7, the droplets either moved toward boundary regions or through the single slit. When moving toward boundaries (fig 19), the droplets often stopped at the far left or far right corners between the boundary and the 'wall' leading to the slit. Droplets also moved toward the shallow boundary above the initial placement region (fig 19(c)). 

\begin{figure*}[b!]
    \begin{subfigure}{0.33\textwidth}
    \includegraphics[width=\linewidth]{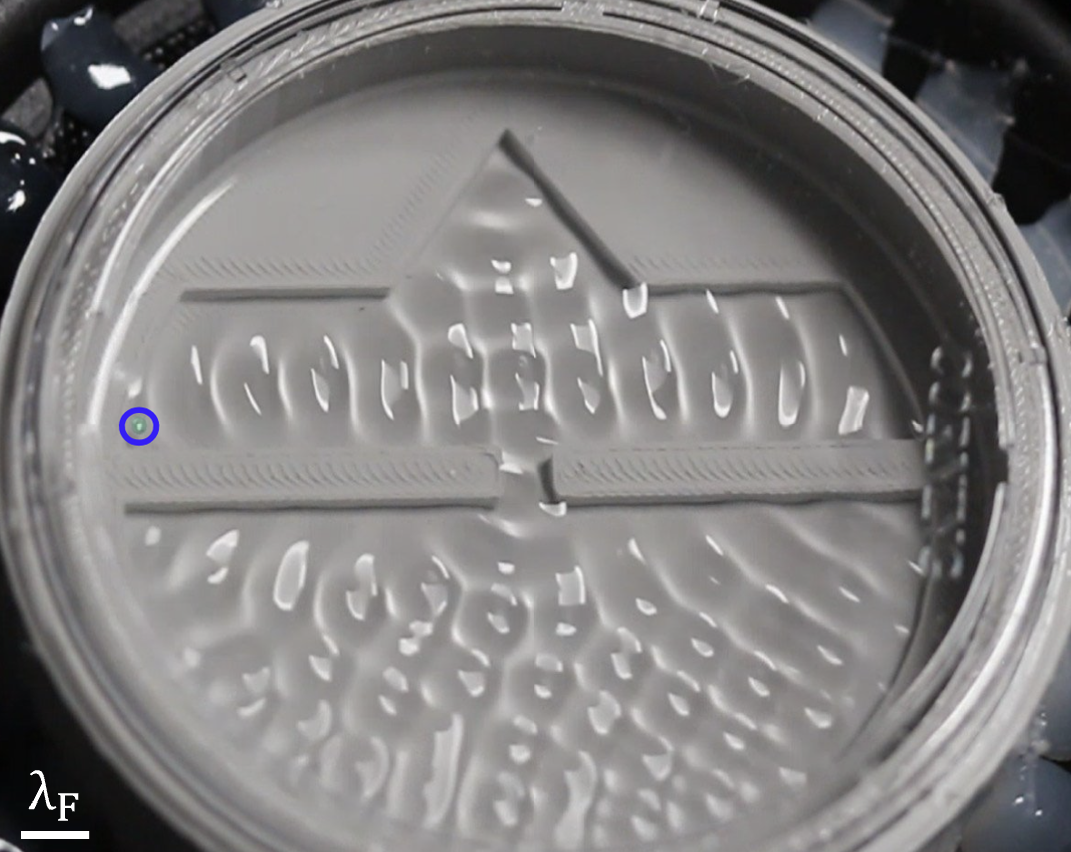}
    \label{fig:1a}\caption{}
    \end{subfigure}%
    \hspace*{\fill}
    \begin{subfigure}{0.33\textwidth}
    \includegraphics[width=\linewidth]
    {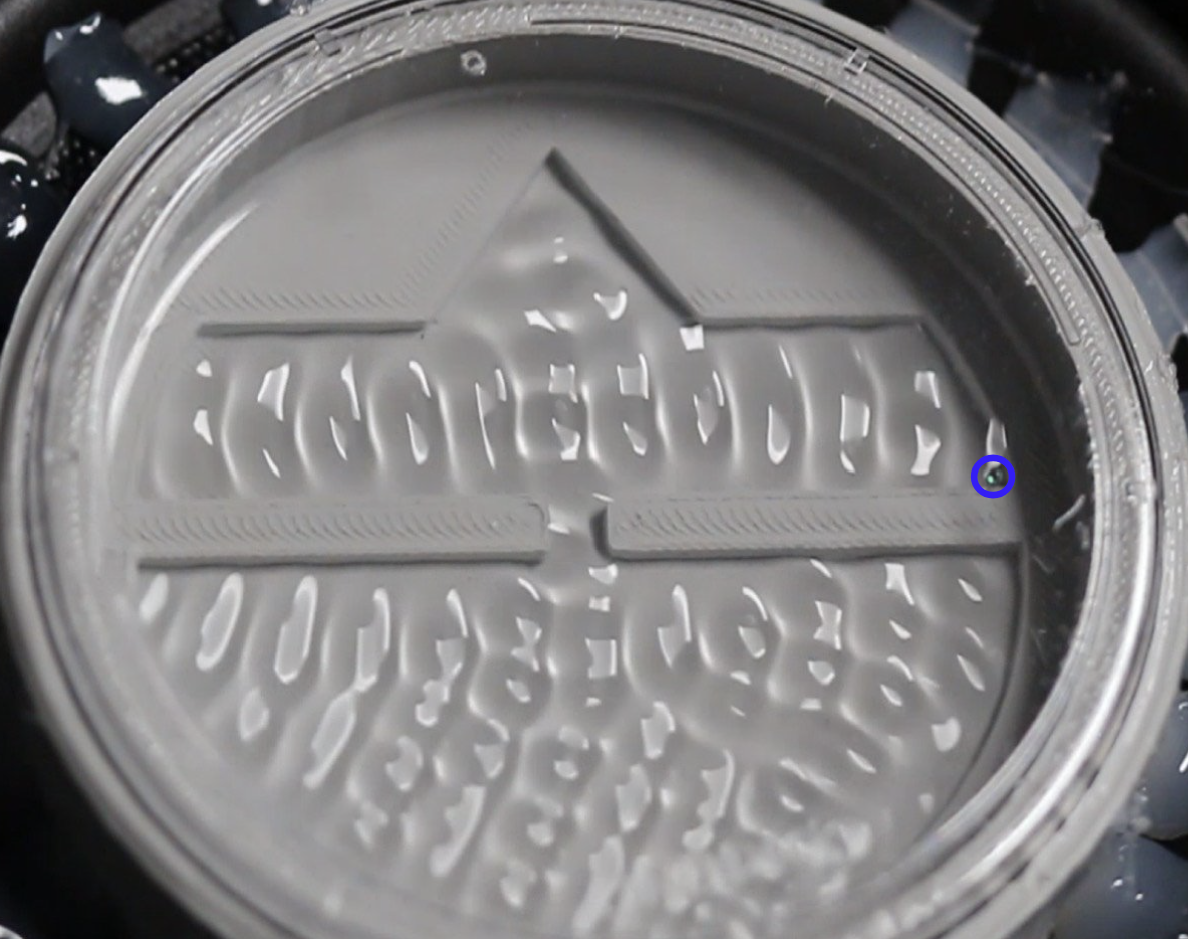}
    \label{fig:1b}\caption{}
    \end{subfigure}%
    \hspace*{\fill}
    \begin{subfigure}{0.33\textwidth}
    \includegraphics[width=\linewidth]
    {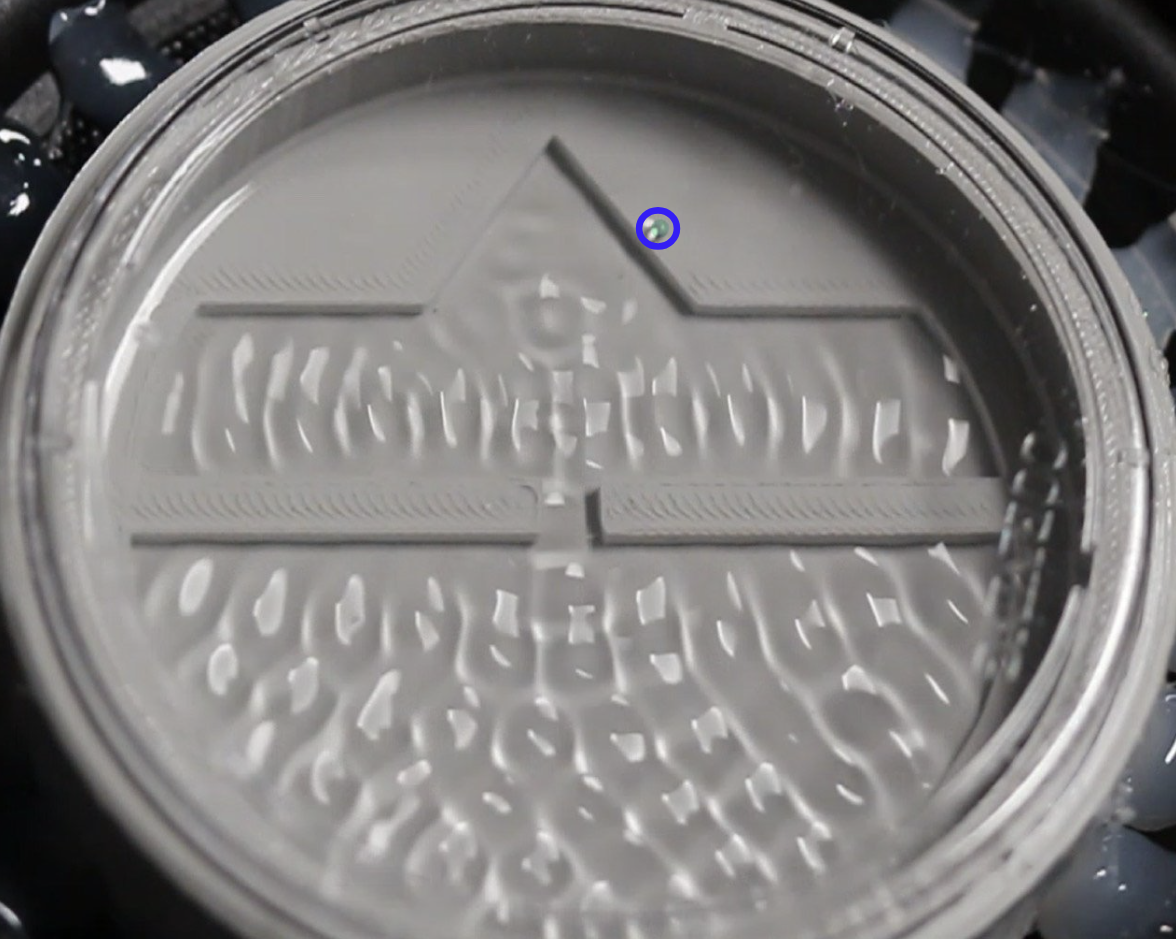}
    \label{fig:1c}\caption{}
    \end{subfigure}%
    \caption{Drops over bath with single-slit indentation and symmetric nonlinear wave topography ($\Gamma \approx$ 7.7). Most drops do not go through the slit, but more often walk (randomly, erratically, or in a straight path) towards specific bath boundaries of shallow depth. In particular, droplets move to the left(a), right(b), or directly above in the y direction (c).}
\end{figure*}
Droplets that passed through the slit exhibited both meandering and erratic movement, including passing through the slit and then changing directions to move back through the slit (fig 20(a)), walking toward boundaries or 'walls'(fig 20(b)), or meandering until coalescence (fig 20(c)).

\begin{figure*}[b!]
    \begin{subfigure}{0.33\textwidth}
    \includegraphics[width=\linewidth]{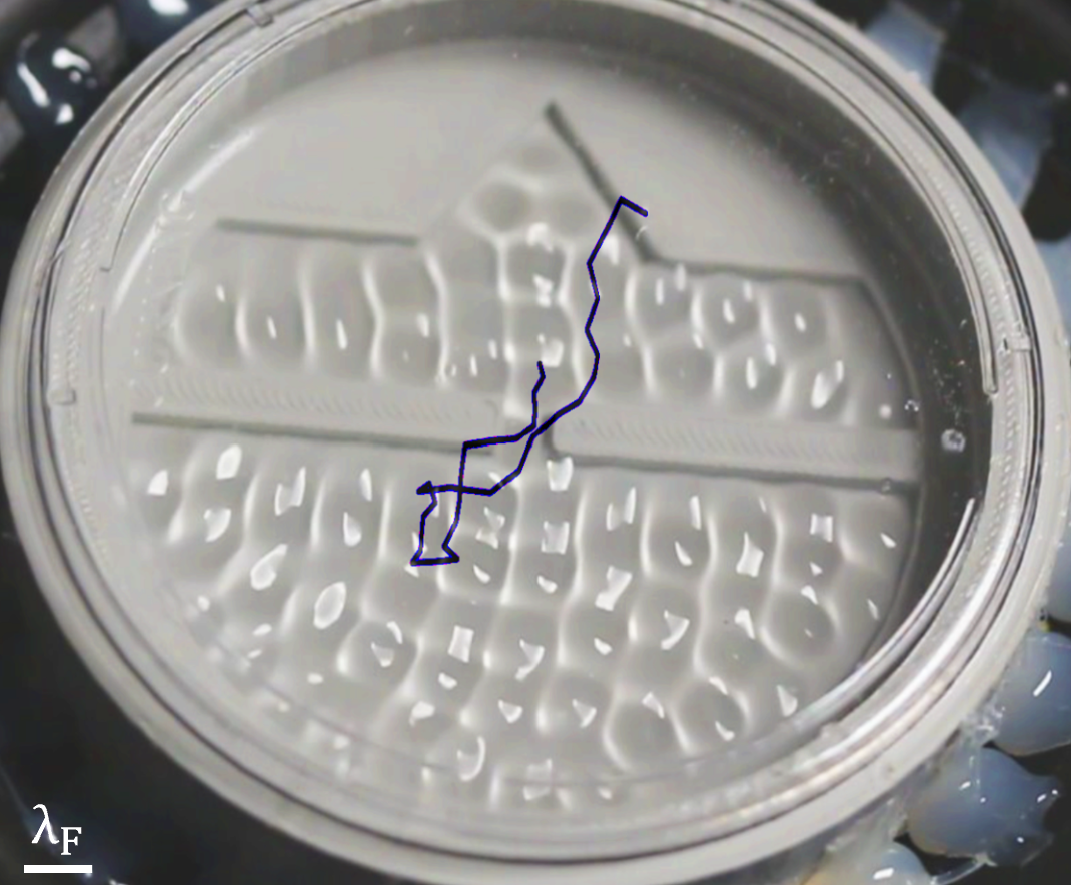}
    \label{fig:1a}\caption{}
    \end{subfigure}%
    \hspace*{\fill}
    \begin{subfigure}{0.33\textwidth}
    \includegraphics[width=\linewidth]{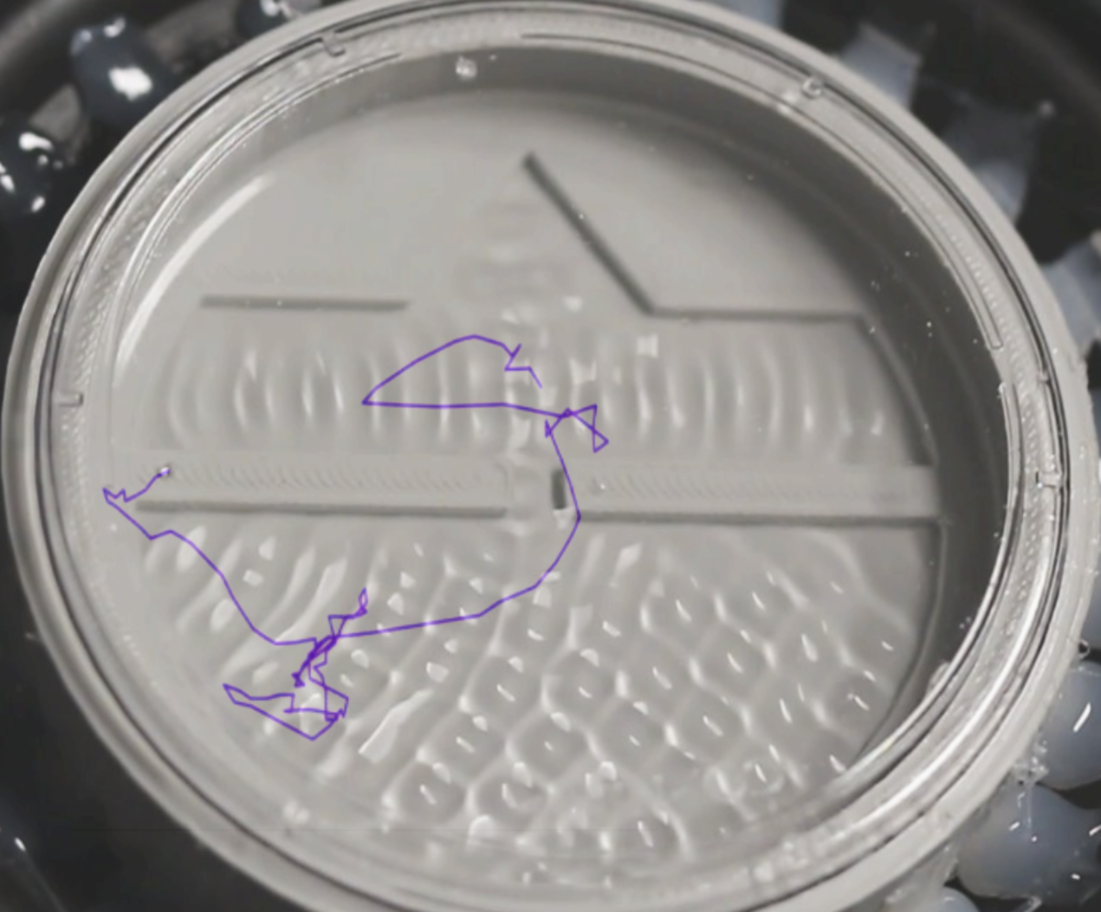}
    \label{fig:1b}\caption{}
    \end{subfigure}%
    \hspace*{\fill}
    \begin{subfigure}{0.33\textwidth}
    \includegraphics[width=\linewidth]{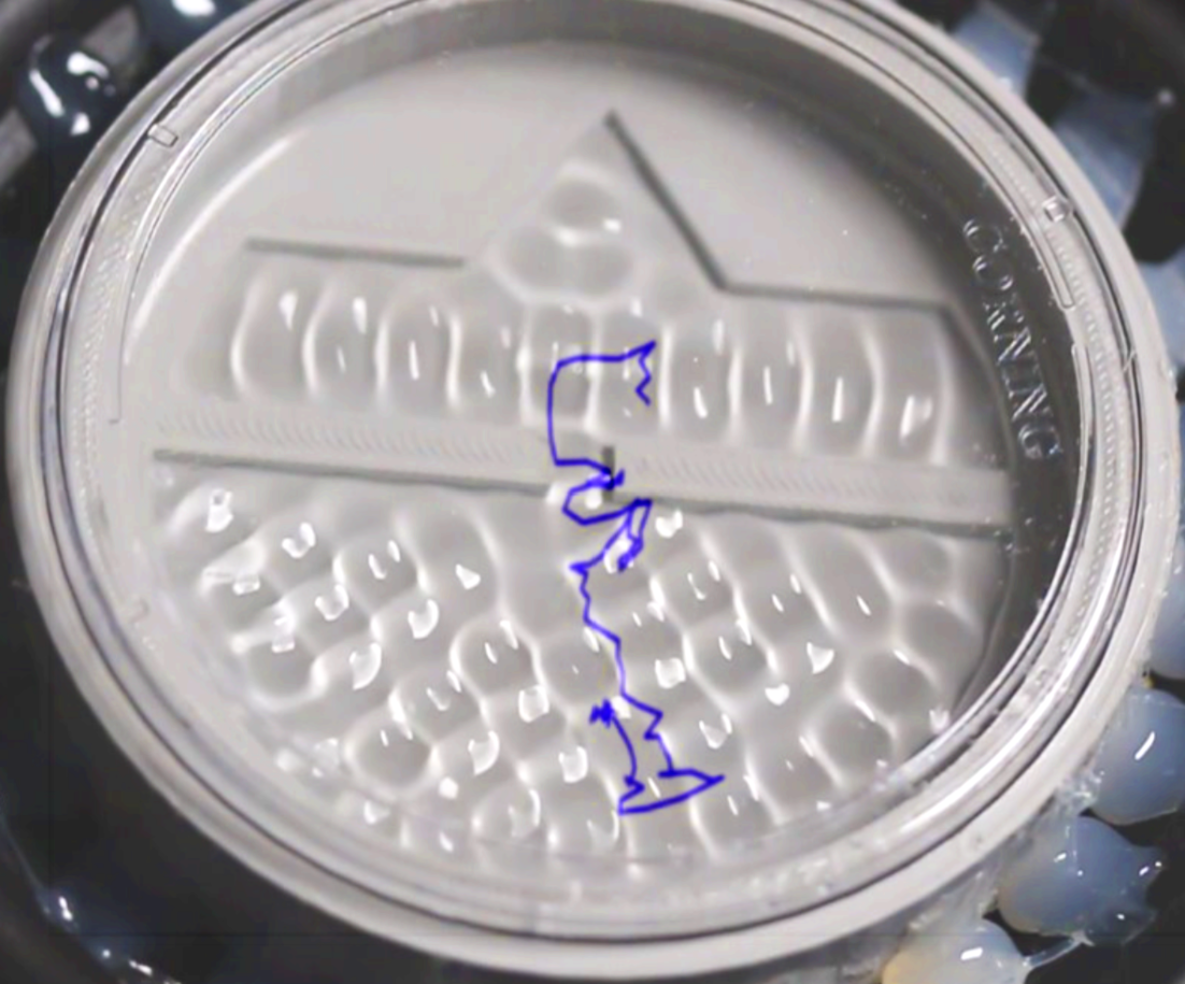}
    \label{fig:1c}\caption{}
    \end{subfigure}%
    \caption{Droplet trajectories over bath with single-slit indentation and symmetric nonlinear wave topography ($\Gamma \approx$ 7.7). Droplet begins just above the slit, passed through the slit, and returned back up through the slit to a 'higher' point (a). Droplet meanders through the slit and to the left side wall, occasionally changing direction erratically both before and after going through the slit (b). Droplet meanders through the single-slit and eventually coalesces is the middle of the bath (c)}
\end{figure*}

\subsection{Droplets in Supercritical Double-Slit Experiments}
In double-slit configurations of an 85 mm inner diameter bath, we observed the same boundary preferences as shown in figure 19. In the double-slit experiments, we found that droplets that walked through one slit could change directions in the bath and move back 'upward' through the other slit (fig 21(a)).

\begin{figure*}[b!]
    \begin{subfigure}{0.49\textwidth}
    \includegraphics[width=\linewidth]{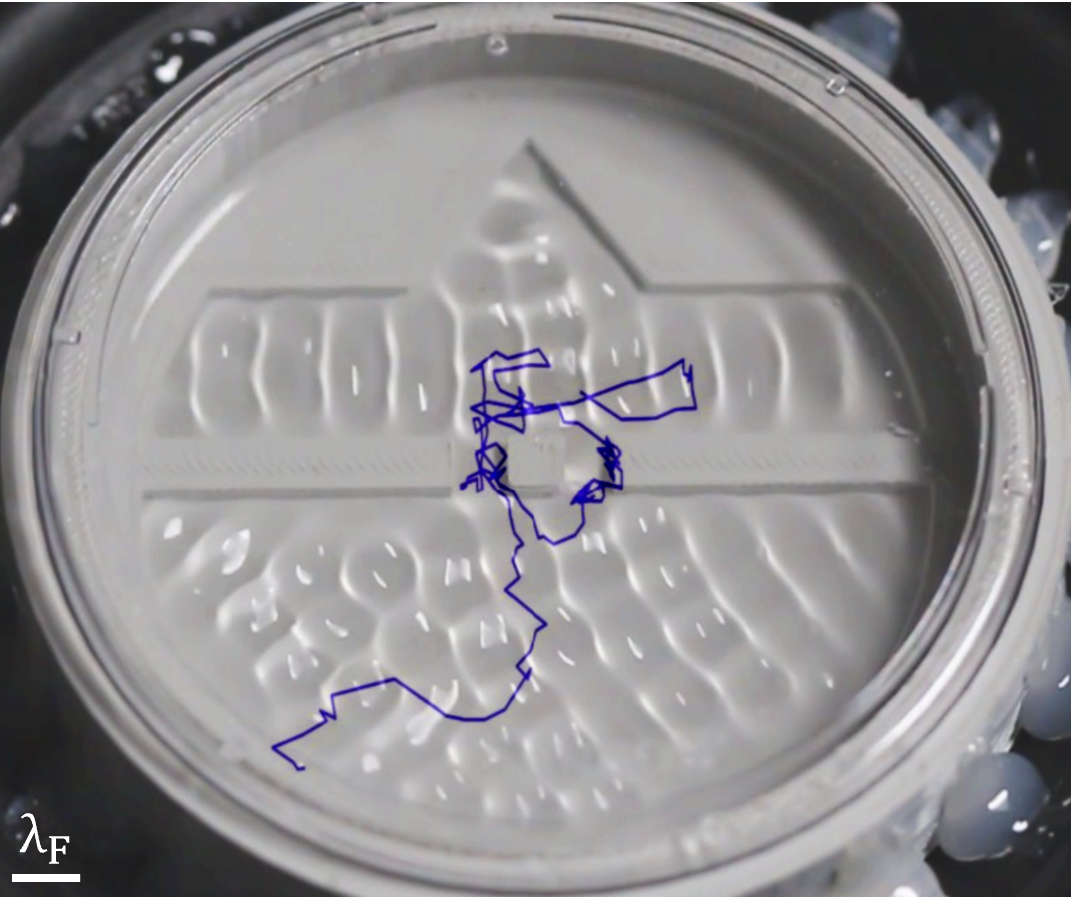}
    \label{fig:1a}\caption{}
    \end{subfigure}%
    \hspace*{\fill}
    \begin{subfigure}{0.49\textwidth}
    \includegraphics[width=\linewidth]{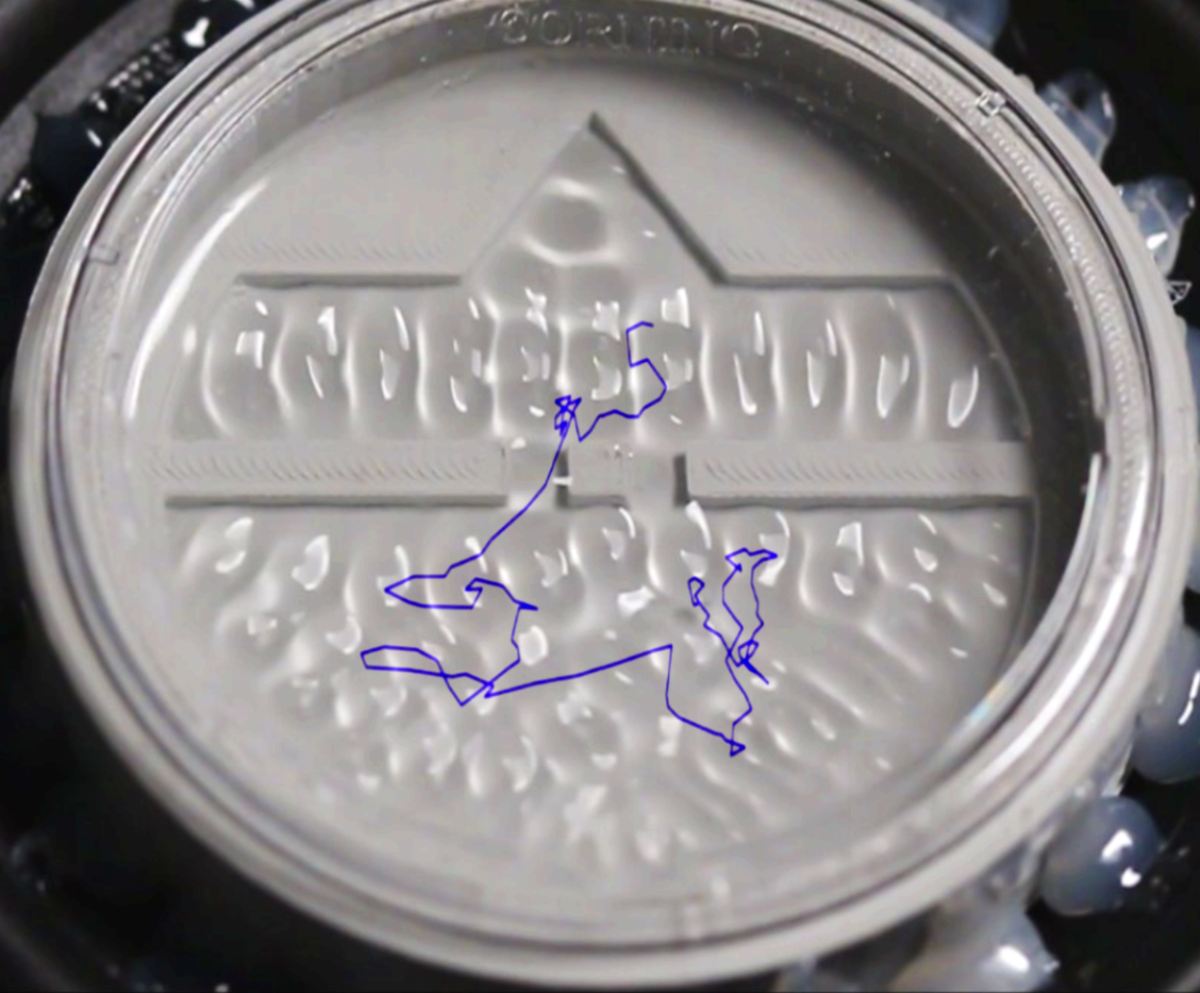}
    \label{fig:1c}\caption{}
    \end{subfigure}%
    \caption{Droplet trajectories over bath with double-slit indentation and symmetric nonlinear wave topography ($\Gamma \approx$ 7.7). Droplet traverses double-slit barrier indentation and meander through the right-side slit, curves its trajectory to go upward through the left-side slit to walk in the lower region of the bath (a). Droplet meanders through the left-side slit and into the lower bath region before coalescing in the middle region (b).}
\end{figure*}

In other 'through slit' instances, as was observed in single slit experiments, droplets walked through slits and meandered in the middle of the bath until coalescence (fig 21(b)).

We found that in the double-slit experiments, many droplets that passed through one of the two slits would walk toward one of the side boundaries (fig 22). Droplets would move to either side boundary regardless of which slit they passed through. The droplets' walking toward these boundaries could be nearly linear walking with near-straight trajectories or display a mix of meandering and erratic movement.

\begin{figure*}[b!]
    \begin{subfigure}{0.33\textwidth}
    \includegraphics[width=\linewidth]{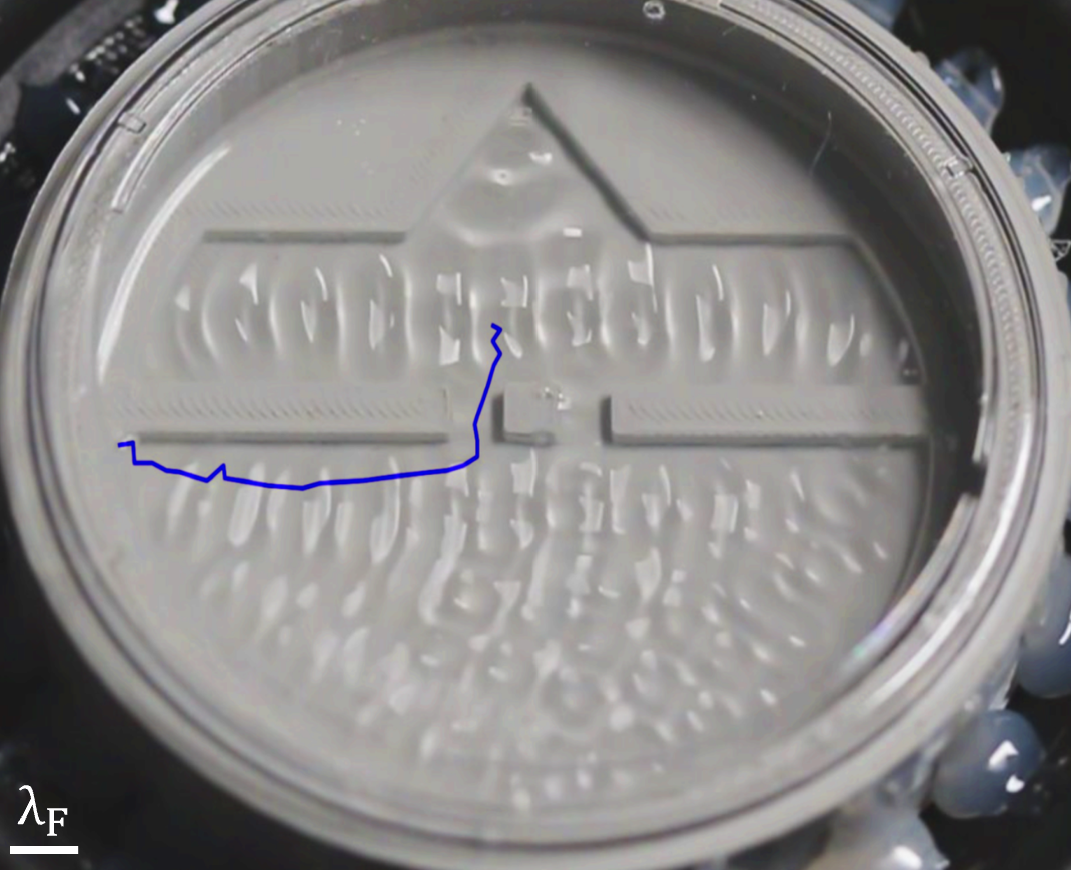}
    \label{fig:1a}\caption{}
    \end{subfigure}%
    \hspace*{\fill}
    \begin{subfigure}{0.33\textwidth}
    \includegraphics[width=\linewidth]{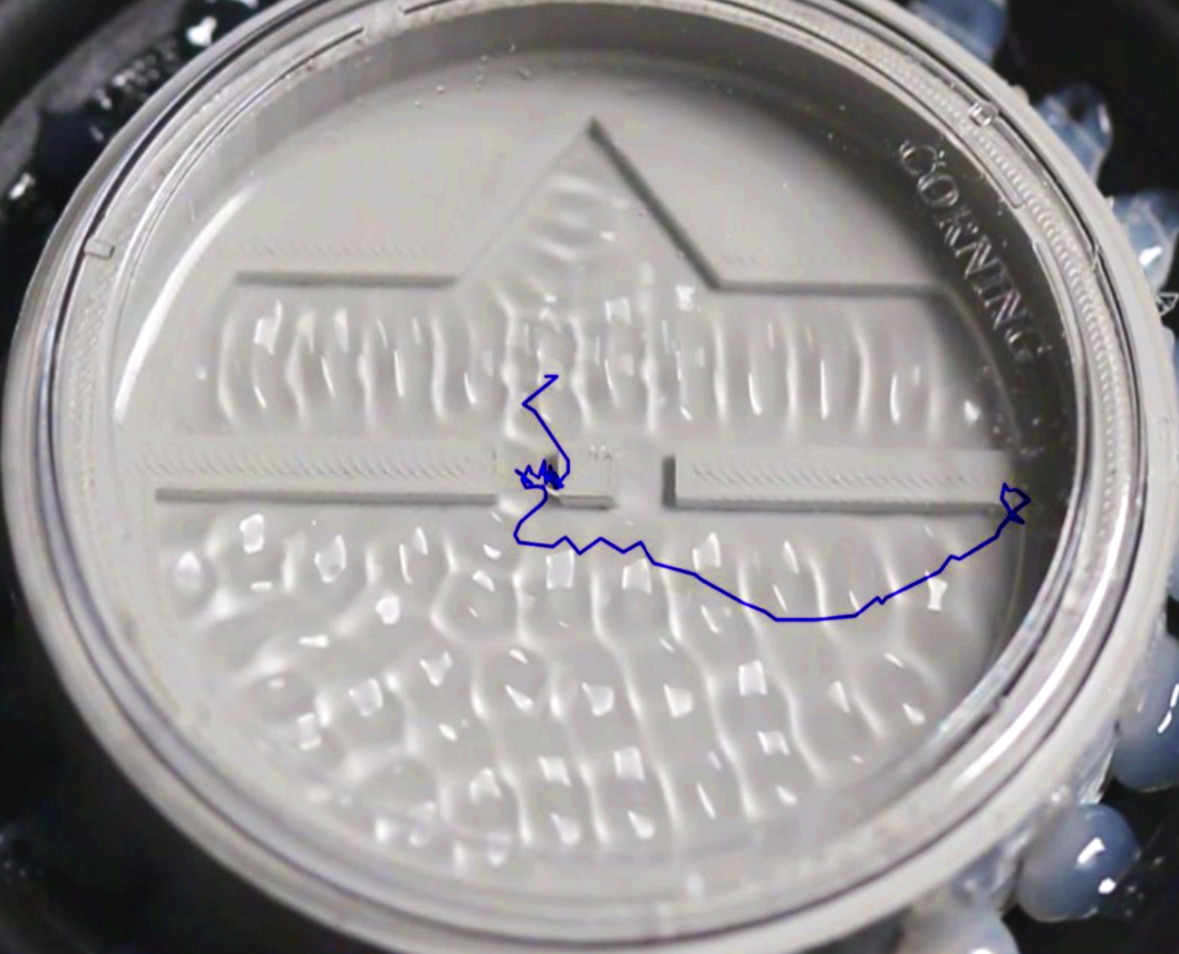}
    \label{fig:1b}\caption{}
    \end{subfigure}%
    \hspace*{\fill}
    \begin{subfigure}{0.33\textwidth}
    \includegraphics[width=\linewidth]{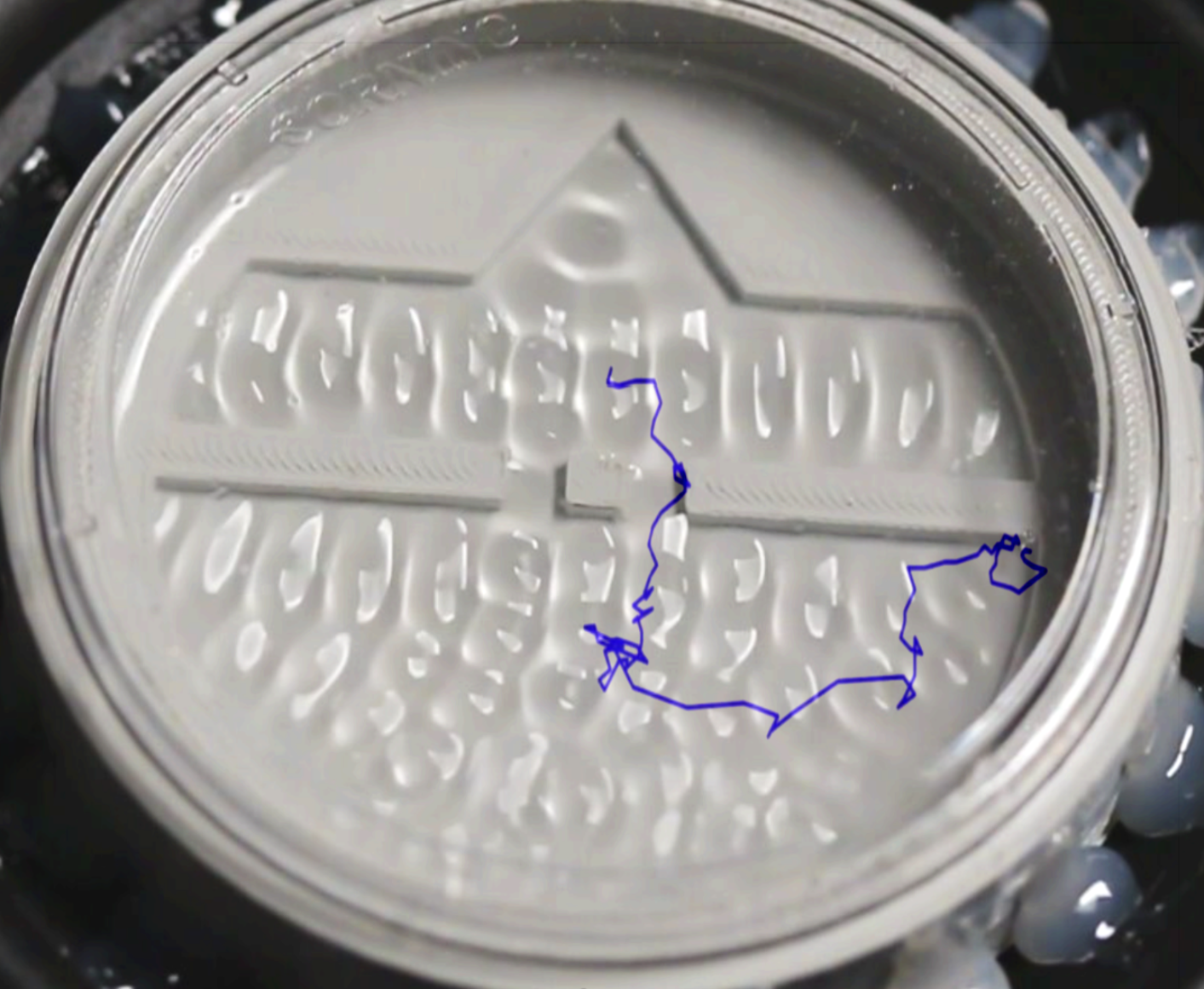}
    \label{fig:1c}\caption{}
    \end{subfigure}%
    \caption{Trajectories of droplets traversing double slit barrier indentation over symmetric nonlinear wave topography ($\Gamma \approx$ 7.7). Droplets here show preference to move towards the top side walls beyond the double slit barrier, sometimes with near-straight trajectories (a-b), and other times meandering in the lower bath region before moving to the side wall (c).}
\end{figure*}
\FloatBarrier
\section{Discussion}
Our results agree with Tambesco et al. \cite{Tamb18}, as we found that droplets bouncing on a bath forced beyond the Faraday acceleration $\Gamma_F$ tend to form clusters and can move in both meandering and erratic trajectories, and that baths forced well beyond $\Gamma_F$ will spontaneously form droplets.

To our knowledge, there has not yet been experiments on walking droplet behavior in a bath of radially uniform topography and spatially nonuniform onset of Faraday waves. We found that droplets move away from the region of initial Faraday pattern onset. Based on the gradient of wave formation, these experiments exhibited an effective gradient of local Faraday acceleration, where droplets preferred regions of lower Faraday acceleration. In regions of higher Faraday acceleration, droplet trajectory is dominated by interactions with Faraday waves, as opposed to a droplet being primarily guided by its own wave field as in lower Faraday accelerations. When in regions forced at or beyond the onset of Faraday patterns, droplets meander and move erratically until forced away from the region. In a region of lesser acceleration, just below the onset of Faraday patterns, droplets are said to be in a 'high memory' regime, in which the radial wave generated by a droplet influences its subsequent trajectory \cite{CastroBush2023}. We observe that the effect of previously generated waves is less prevalent in lower effective Faraday accelerations, causing the droplet to walk toward regions of lesser and lesser Faraday acceleration until they are below a 'walking' acceleration threshold. The initial movement away from the region of higher effective Faraday acceleration is likely due to this region behaving as a localized source of vertical acceleration as compared with the larger bath. As stated in the previous section, we have not determined a definitive cause for this spatially nonuniform onset of Faraday patterns.

In our single- and double-slit experiments for walking droplets in the supercritical Faraday regime, unlike the subcritical quantum slit analog experiments of Pucci et al., we did not find that trajectories move outward radially from the slit iand directly toward the bath edge \cite{Pucci18}. Nor did we find any point beyond the single- or double-slit apparatus in which we could predict the subsequent trajectory of the droplet. Due to the influence of the Faraday waves on droplet trajectory, we observed chaotic directional changes in stark contrast to the subcritical case. 

We note that our methods of drop placement differed from Pucci et al., in that we did not use a specialized drop launcher. Thus, in our experiments, droplets were free to move upward from the region of initial drop placement and allowing a larger maximum deflection angle than that of the Pucci et al experiments. In our double-slit experiments, we found that droplets moved toward the far boundaries once past slits. This may be an influence of the 'edge' of the slits, similar to that observed by Pucci et al., or may be attributed to the droplets observed preference away from visibly turbulent regions.
The methods of our experiments lend to a very low portion of droplets passing through the slits, so a statistical distribution of droplets through slits was not determined. While differing from the behavior of quantum single- and double-slit experimental results as droplets move back through slits, the unpredictable trajectory of droplets passing through slits and meandering in the bath is conceptually similar to the unknowable location of quantum particles before observation. We believe this warrants further investigation with a revised experimental setup in order to determine a statistical distribution of single- and double-slit trajectories in the supercritical regime. A complication in creating such a distribution would use determining the end point of trajectories, as droplets may move toward boundaries and stop horizontal motion within boundary regions, or continue to meander within the middle of the bath until they coalesce with the fluid bath.
\FloatBarrier
\section{Conclusion}
\FloatBarrier
We have presented an inexpensive, relatively simple apparatus for walking droplet experiments in the supercritical Faraday acceleration regime. We executed novel experiments in this supercritical regime on single- and double-slit bath configurations. We documented a case of asymmetric effective Faraday acceleration for radially uniform bath topography and found that droplets move away from regions of greater effective Faraday acceleration. In our single- and double-slit experiments, we found that Faraday waves dominate droplet trajectory and cause droplets to meander and move erratically before and beyond the slits, and discussed how this unpredictable behavior is conceptually similar to uncertainty in the locations and momentum of quantum particles. We found that droplets often move towards points with the highest deflection angle from the slit, and that this may motivate future researchers to investigate a similar experimental apparatus with the goal of creating a statistical distribution of walking droplets in supercritical single- and double-slit regimes to confirm this bias.

\begin{acknowledgments}
We would like to thank Dr. Henry Greenside of the Duke Physics Department for stimulating discussions on nonlinear pattern formation.
\end{acknowledgments}

\section*{Author Declaration}
The authors have no conflicts to disclose.

\section*{Data Availability Statement}
Data will be made available by the corresponding author upon request via email.

\appendix

\section{Hydrodynamic Quantum Theory}

In recent years, the Hydrodynamic Quantum Field Theory (HQFT) has emerged as a new perspective on de Broglie-Bohm pilot wave quantum mechanics as informed by the dynamical behavior exhibited in select walking-droplet experiments. Such experiments claim to convey behavior analogous to quantum phenomena, such as wave-particle duality, quantized states, single-particle diffraction and interference, and tunneling \cite{CastroBush2023}. These analogies are made through following the dynamics emergent from pilot wave quantum mechanics, in which particles are guided by an intrinsic vibration that generates a pilot wave, whose wave field guides the motion of the particle \cite{Dagan20}. Where de Broglie left the nature of wave generation by the particle unspecified, HQFT asserts that capillary Faraday waves act as an analogy to de Broglie matter waves, and the droplet bouncing on the wave bath is analogous to said wave generation by the particle \cite{Dagan20}. This analogous behavior is primarily studied in droplets 'walking' over baths forced at subcritical Faraday accelerations \cite{CastroBush2023}, as at the onset of Faraday patterns, the droplet is no longer primarily guided by its own wavefield, but becomes heavily influenced by the strong Faraday waves, differing from the description of particle dynamics suggested in HQFT. 
In de Broglie-Bohm mechanics, theorists have discussed the role of chaos in causing quantum systems to relax towards a "quantum equilibrium" in which their behavior approaches satisfaction of Born's rule for quantum probabilities (probability $p = \left| \psi \right |^2$) \cite{Efth17}. For generic systems with an initial deviation from Born's rule, chaos was deemed essential in relaxing the behavior of these systems towards satisfying $p = \left | \psi \right | ^2$, as ordered trajectories alone would remain in deviation \cite{Efth17}. 
We note that chaos in this sense is due to the geometry of a system's wavefunction (a solution of the linear Schrodinger equation), differing from the chaos exhibited in erratic or meandering walking-droplet trajectories or spatiotemporal chaos in Faraday wave patterns due to nonlinear dynamical interactions. This difference speaks to a specific limitation in our own experiments of the single- and double-slit hydrodynamic quantum analogs, as the chaotic trajectories we produced arose from the influence of nonlinear Faraday wave patterns. However, we believe that as chaos is a known accelerant of de Broglie-Bohm quantum systems toward $p = \left | \psi \right | ^2$ behavior, it is desirable to explore how chaos influences the distribution of droplet positions in experiments such as the single- and double-slit hydrodynamic analog in which ordered trajectories have not yet produced quantum-like diffraction patterns.

\section{Visualization and Tracking Methods}
We tested two open-source computer vision software for droplet tracking applications, Google Deepmind's Track Any Point (TAP) software and Meta AI's Segment Anything Model (SAM).
In the TAP software, we attempted to track points in a simple simulation of a red circle bouncing on a standing wave (fig 25). 
\begin{figure}
    \centering
    \includegraphics[width=1.0\linewidth]{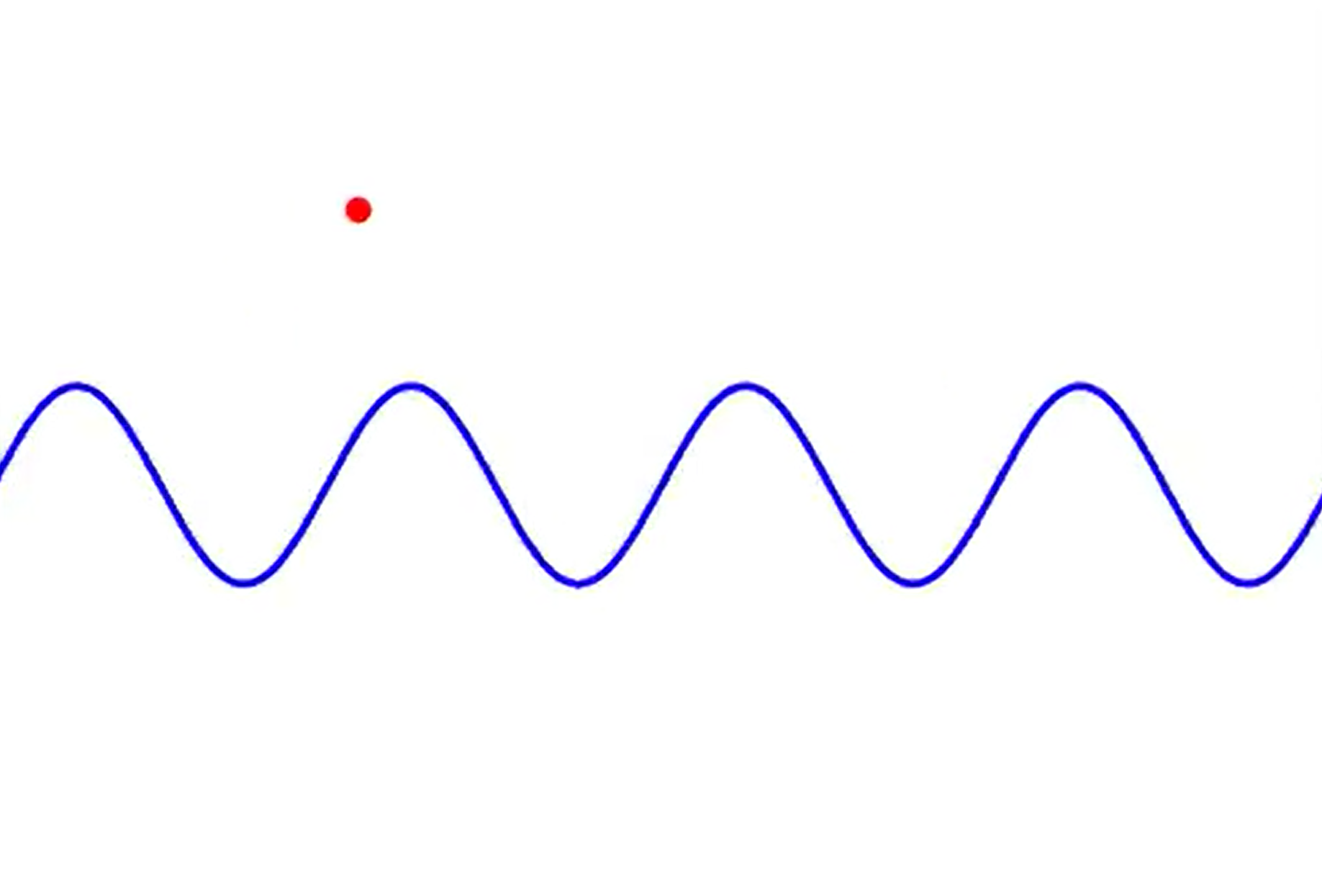}
    \caption{Image of red circle bouncing on a standing wave, a simplified model of the walking droplet system for computer vision software testing.}
    \label{fig:placeholder}
\end{figure}

We found that TAP software recognized the red circle until the circle collided with the standing wave. In testing the TAP software on real images of our experiment, the software was ineffective and unreliable in tracking droplets, as droplets moved erratically and Faraday wave patterns occasionally hindered droplet visibility. Ultimately, we concluded that TAP software is not yet capable of effectively tracking walking droplets in the supercritical Faraday acceleration regime.
When using SAM, we attempted to mask droplets and use the location and shape of a droplet mask in an initial frame (fig 26) to identify drop location in subsequent frames.

\begin{figure}
    \centering
    \includegraphics[width=1.0\linewidth]{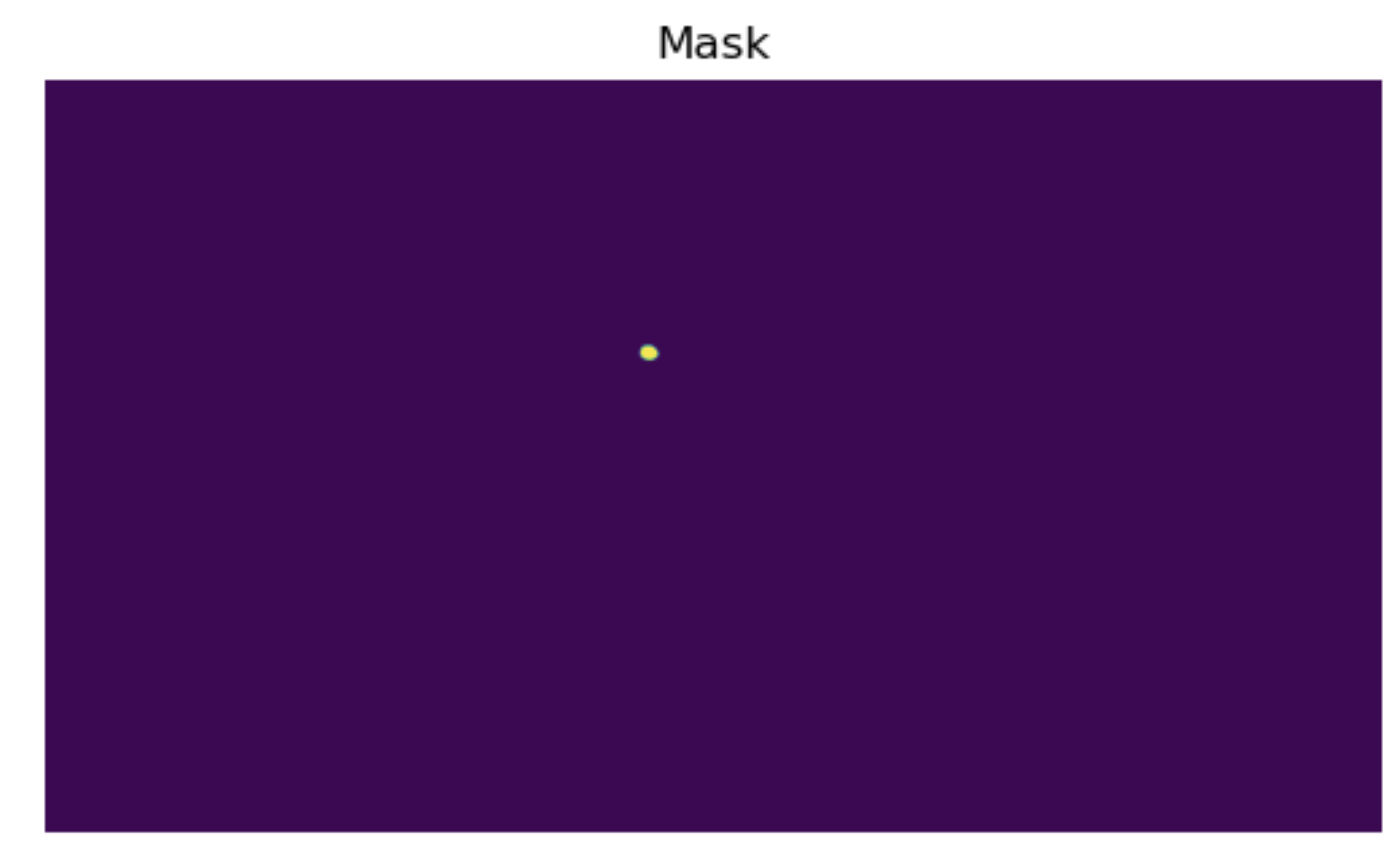}
    \caption{Mask of droplet from experimental data created using SAM software.}
    \label{fig:placeholder}
\end{figure}

We found that while capable of producing masks of droplets, the SAM software did not create a mask of the walking droplet in every frame where a droplet was present. Additionally, in our use of SAM software, we found that other masks were occasionally created for Faraday waves and sections of the bath geometry that met the same location and size requirements of the droplet, making it difficult to track a droplet using this method.

Ultimately, we used the Vernier's Video Physics mobile app and graphical analysis computer software to manually track droplets by selecting droplet location in each video frame. This method produced the most consistent and accurate trajectory graphs and is an accessible method for reproduction of our experiment or similar experiments.

\subsection{References}
\nocite{*}
\bibliography{your-bib-file}

\end{document}